\documentclass[natbib,smallextended]{svjour3}

\usepackage{amsmath,dsfont,natbib,graphicx,lineno}
\usepackage{epsfig}
\usepackage{epstopdf}
\usepackage{graphicx}
\usepackage{psfrag}

\usepackage{amssymb,subfigure}
\usepackage{mathrsfs}
\usepackage{color}

\def\R{\mathbb R}

\def\lp{\left(}
\def\rp{\right)}

\def\epsilon{\varepsilon}
\def\ds{\displaystyle}
\def\lp {\left( }
\def\rp {\right) }

\renewcommand{\thesubsection}{\arabic{section}.\arabic{subsection}}
\renewcommand{\thesection}{\arabic{section}}

\newcommand{\be}{\begin{equation}}
\newcommand{\ee}{\end{equation}}
\newcommand{\baa}{\begin{array}}
\newcommand{\eaa}{\end{array}}
\newcommand{\ba}{\begin{eqnarray}}
\newcommand{\ea}{\end{eqnarray}}
\newcommand{\bi}{\begin{itemize}}
\newcommand{\ei}{\end{itemize}}

\date{}

\begin{document}

\title{Using genetic data to estimate diffusion rates in heterogeneous landscapes\thanks{The research leading to these results has received funding from the French Agence Nationale pour la Recherche, within the ANR-12-AGRO-0006 PEERLESS, ANR-13-ADAP-0006 MECC and ANR-14-CE25-0013 NONLOCAL projects and from the European Research Council under the European Union's Seventh Framework Programme (FP/2007-2013) / ERC Grant Agreement n.321186 - ReaDi - Reaction-Diffusion Equations, Propagation and Modelling.}}

\author{L Roques         \and E Walker \and P Franck \and
        S~Soubeyrand \and E K Klein
}


\institute{L Roques \at
              INRA, UR 546 Biostatistique et Processus Spatiaux, 84000 Avignon, France \\
              \email{lionel.roques@avignon.inra.fr}           
           \and
            E Walker \at
            INRA, UR 546 Biostatistique et Processus Spatiaux, INRA, 84000 Avignon, France
            \and
            P Franck \at
             INRA, UR 1115 Plantes et Syst\`emes de Culture Horticoles,  84000 Avignon, France
             \and
            S Soubeyrand \at
            INRA, UR 546 Biostatistique et Processus Spatiaux, INRA, 84000 Avignon, France
            \and
            E K Klein \at
            INRA, UR 546 Biostatistique et Processus Spatiaux, INRA, 84000 Avignon, France
}

\date{Received: date / Accepted: date}

\maketitle


\begin{abstract}

Having a precise knowledge of the dispersal ability of a population in a heterogeneous environment is of critical importance in agroecology and conservation biology as it can provide management tools to limit the effects of pests or to increase the survival of endangered species. In this paper, we propose a mechanistic-statistical method to estimate space-dependent diffusion parameters of spatially-explicit models based on stochastic differential equations, using genetic data.
Dividing the total population into subpopulations corresponding to different habitat patches with known allele frequencies, the expected proportions of individuals from each subpopulation at each position is computed by solving a system of reaction-diffusion equations. Modelling the capture and genotyping of the individuals with a statistical approach, we derive a numerically tractable formula for the likelihood function associated with the diffusion parameters.

In a simulated environment made of three types of regions, each associated with a different diffusion coefficient, we successfully estimate the diffusion parameters with a maximum-likelihood approach.  Although higher genetic differentiation among subpopulations leads to more accurate estimations, once a certain level of differentiation has been reached, the finite size of the genotyped population becomes the limiting factor for accurate estimation.
\end{abstract}

\section{Introduction \label{sec:Intro}}

Dispersal is one of the main forces driving population redistribution \citep{Tur98}, gene flow \citep{Sta87,Boh99} and genetic diversity \citep{Hew00,RoqGar12,RoqHos14}.  Dispersal
directly affects population flows between different spatial positions. Having a precise knowledge of these flows is of critical importance in agroecology and conservation biology, as it can provide management tools to limit the effects of pests \citep{Gil08,PapGoy11} or to increase the survival of endangered species \citep{HanGil96} by acting on the landscape structure.

In heterogeneous environments, dispersal is often influenced by local landscape features. This local effect of the landscape on mobility can be captured by most spatially-explicit models at the scale of individual movement, such as in random walk models and stochastic differential equations \citep{PreAge04,SmoFoc10} or at the scale of population density in reaction-diffusion models \citep{Shikaw97,CanCos03,OvaRek08,Roq13}, through a space-dependent mobility parameter.

Traditionally, Euclidian distances and least-cost distances have been used, e.g. in models of isolation by distance \citep{Wri43,Rou97,BroRay06}, to quantify the population flows.  However, these approaches have several drawbacks; in particular, least-cost distances are based on a subjective definition of a cost function and assume a single and optimal migration path. Resistance-based approaches are more realistic \citep{McR06,GraCha14}, as the resistance distance is computed in a similar manner as in an electrical network, where all possible paths are taken into account.  However, the computation of resistance distances is time-consuming and generally does not allow for a precise estimation of the local resistance parameter, e.g. by maximum likelihood, but rather to test a limited set of conjectured resistance values \citep{GraBei13}, leading again to a subjective parametrization of the local resistance values.
Additionally, although there exist random walk interpretations of the effective resistance in electrical networks \citep{DoySne84}, these approaches are not based on a mechanistic description of the spatio-temporal dynamics of a population \citep[see Theorems 1 and 2 in][]{Tet91}, and therefore do not directly quantify population flows.

The method developed in this paper enables direct and fast estimations of a spatially-heterogeneous parameter $D(x)$ measuring the local mobility of individuals at the space position $x,$ in mechanistic models based on stochastic differential equations. More precisely, we assumed that the individual trajectories followed It\={o} diffusion processes, corresponding to uncorrelated random walks with spatially-varying speed. This framework is widely used for analyzing movement, see~\cite{PreAge04,SmoFoc10} and the references therein. In this framework, the expected population density at any time and space position can be computed using the corresponding Fokker-Planck partial differential equation \citep{Gar09} with diffusion parameters $D(x)$. Here, we also took into account death events occurring at exponentially distributed times, leading to a reaction-diffusion description of the population density. Thanks to a well-developed theory for their numerical analysis and efficient softwares (e.g., freefem++, see \cite{Freefem12}, or Comsol Multiphysics$^\copyright$), the numerical computation of the solutions of such reaction-diffusion equations is fast and reliable, even in the presence of heterogeneous coefficients. This makes them ideally suited for parameter estimation \citep{SouRoq14}.

Several types of data can be used for the estimation of dispersal parameters in population models. Most studies bear on abundance data \citep{RoqSouRou11} or on mark-recapture experiments \citep{Tur98,OvaRek08}, where individuals are marked with different technics, such as color markers or radioactive isotopes \citep{SouHen09}.
As opposed to this marking experiments, passive surveys of the relative abundance of genetic markers that are naturally present in a population lead to spatio-temporal data which can be easier to obtain and more informative \citep{Rob12}. Here, we considered the dispersal of individuals starting from several habitats and which eventually died, defining the end of the dispersal period. The estimation of the parameter $D(x)$ was based on measurements of the genotypes of individuals captured at several positions during the dispersal process, the location of the habitats and the allele frequencies before dispersal in these sites being considered as known. Although numerous methods have been developed in landscape genetics to estimate dispersal from molecular markers \citep{HamTra11} we are not aware of any article estimating diffusion parameters from such data.

The method is based on a mechanistic-statistical approach \citep{Ber03,Wik03,OvaRek08,SouLai09,RoqSouRou11,SouRoq14} included in the framework of state-space models \citep{PatTho08,DurKoo12}. This  approach typically combines a mechanistic model describing the dynamics under investigation with a statistical model conditional on the dynamics, describing how the measurements have been collected, bridging the gap between the data and the model for the dynamics. In the mechanistic part of our model, we divided the total population into subpopulations at Hardy-Weinberg equilibrium, each one corresponding to the individuals coming from a different habitat patch. The dynamics of the different subpopulations were then described by a system of reaction-diffusion equations, as in \cite{RoqGar12}. Given a diffusion coefficient $D(x)$, this allowed us to compute the  expected proportions of individuals from  each subpopulation at each space position. Conversely, the genotype data contain information about these proportions; namely, the probability to observe a given genotype at some trapping location depends on the respective contributions of each subpopulation and on the allele frequencies in these subpopulations. Modelling the capture of the individuals with a statistical approach, and using the Hardy-Weinberg equilibrium assumption in each subpopulation as a key ingredient, it was therefore possible to derive a numerically tractable formula for the likelihood function associated with the diffusion parameters $D(x),$ given the genotypes of the captured individuals.


\noindent \textit{Note: a summary of the notations used throughout this paper is provided in Table~\ref{tab:notations}.}

\begin{table}[h!]
   \centering
\begin{tabular}{|c|c|}
    \hline
    Notation & Explanation \tabularnewline
    \hline
    $u(t,x)$ & density of  dispersers \tabularnewline\hline
    $u^h(t,x)$ & density of dispersers coming from habitat $\Omega^h$  \tabularnewline\hline
    $u_0(x)$ & pre-dispersal density\tabularnewline\hline
    $u_0^h(x)$ & pre-dispersal density  of individuals  coming from habitat $\Omega^h$ \tabularnewline\hline
    $\alpha$ & pre-dispersal density  in the habitats \tabularnewline\hline
    $w_\infty(x)$ & cumulated density of dispersers \tabularnewline\hline
    $w_\infty^h(x)$ & cumulated density of dispersers  coming from habitat $\Omega^h$\tabularnewline\hline
    $\beta_\tau$ & capture rate in trap $\theta_\tau$ \tabularnewline\hline
    $C_\tau$ & expected number of individuals captured in trap $\theta_\tau$  \tabularnewline\hline
    $C^h_\tau$ & expected number of individuals  coming from habitat $\Omega^h$ \tabularnewline
    &  captured in trap $\theta_\tau$  \tabularnewline\hline
    $D(x)$ and $D_1, \, D_2, \, D_3$& diffusion parameters (mobility) \tabularnewline\hline
    $\nu$ & life expectancy of the dispersers \tabularnewline\hline
    $\Omega$ & study site   \tabularnewline\hline
    $\Omega^h, \, h=1,\ldots,H$ & habitats (subsets of $\Omega$) \tabularnewline\hline
    $\theta_\tau, \, \tau=1,\ldots, J$ & traps (subsets of $\Omega$) \tabularnewline\hline
    $x_h$ & position of the center of the habitat $\Omega^h$ \tabularnewline\hline
    $x_\tau$ & position of the center of the trap $\theta_\tau$ \tabularnewline\hline
    $G$ & number of individuals genotyped in each trap \tabularnewline\hline
    $\lambda=1,\ldots, \Lambda$ & index for the loci  \tabularnewline\hline
    $a=1,\ldots, A_\lambda$ & index for the alleles \tabularnewline\hline
    $(a^1,a^2)$ & couple of alleles at a given locus \tabularnewline\hline
    $p_{h\lambda a}$ & frequency of allele $a$ of locus $\lambda$ in habitat $\Omega^h$ \tabularnewline\hline
    $\mathcal{F}_{h\lambda}$ & allele frequencies at locus $\lambda$ in habitat $\Omega^h$ \tabularnewline\hline
    $\mathcal{G}_{i\tau}$ & genotype of the $i^{\hbox{th}}$ genotyped individual in trap $\theta_\tau$ \tabularnewline\hline
    $\mathcal{M}$ & measurement set consisting of all the genotypes in all traps \tabularnewline\hline

 \end{tabular}
 \caption{Summary of the notations used in the main text.}
 \label{tab:notations}
 \end{table}

\section{Models \label{sec:RD}}

\noindent{\it Modelling dispersal and death.} We begin with a Lagrangian description of the individual movements. We assumed that the positions of the individuals followed 2-dimensional space-\linebreak heterogeneous It\={o} diffusion processes, corresponding to uncorrelated random walks. This means that the individuals travel at random, with no drift in any particular direction \citep{PreAge04,SmoFoc10}. However, the mobility of the individuals can be influenced by their position. The corresponding stochastic differential equation for the position $X_t\in \R^2$ of an individual at time $t$ can be written:\be \label{eq:SDE1}dX_t=\sqrt{2 D(X_t)}dW_t,\ee where $W_t$ is the $2-$dimensional Wiener process (Brownian motion). The coefficient $D(X_t)$ is called the diffusion coefficient. With this model, in a small time interval of length $\tau,$ each coordinate of $X_t$ is incremented by a normally distributed value with mean $0$ and variance $2\,\tau\, D$. Thus, $D(x)$ is a measure of the local mobility of the individuals. When $D$ is constant, the stochastic differential equation \eqref{eq:SDE1} corresponds to the standard Brownian motion.

We also assumed that each individual had a life expectancy $\nu>0,$ the death events being independent and identically distributed and modelled by exponential distributions with parameter $1/\nu$.

Under these assumptions, we can switch to an Eulerian description of the population. The expected  population density $u(t,x)$ at time $t$ and position $x$, starting from an initial distribution $u_0$ satisfies the following Fokker-Planck reaction-diffusion equation \citep[see e.g.][]{Gar09}:
\be\label{eq:RDu} \left\{ \baa{l}\ds \frac{\partial u}{\partial t}=\Delta (D(x) \, u) -\frac{u}{\nu}, \ t>0, \, x\in \Omega, \\
u(0,x)=u_0(x),
\eaa \right.\ee
where $\Delta=\frac{\partial^2 }{\partial x_1^2}+\frac{\partial^2 }{\partial x_2^2}$ is the 2-dimensional Laplace diffusion operator. The set $\Omega\subset \R^2$ is the study region. For this equation to be well-posed, some conditions on the boundary $\partial \Omega$ of $\Omega$ have to be specified. In the computations of this paper, we assumed absorbing conditions ($u(t,x)=0$ on $\partial \Omega$). Reflecting conditions ($\nabla (D(x) u(t,x))\cdot \mathbf{n}(x)=0$ on $\partial \Omega,$ where $\mathbf{n}(x)$ is the outward normal to $\Omega$) could have been assumed as well.

The quantity $u_0$ is called the \emph{pre-dispersal density}. It corresponds to the density of individuals at a motionless stage, such as eggs, pupae or larvas in insect populations or immature seeds in plants.  The quantity $u(t,x)$ is the (expected) \emph{density of dispersers}, e.g. insects at the adult stage or dispersing seeds. We assumed a zero pre-dispersal density outside some known disjoint subsets (habitats) $\Omega^h,$ $h=1,\ldots,H,$ where $u_0$ is positive and constant (equal to $\alpha>0$):
\be
u_0(x)= \sum\limits_{h=1}^{H} \alpha \, \mathds{1}_{x\in \Omega^h}, \hbox{ for all }x\in \Omega,
\ee
where $\mathds{1}_{x\in \Omega^h}$ is the characteristic function of the set $\Omega^h:$ it takes the value $1$ in $\Omega^h$ and $0$ anywhere else. A possible gradual release of the pre-dispersal populations could also be assumed by considering a slightly modified version of the equation \eqref{eq:RDu}, see Appendix~A.

\

\noindent{\it Modelling the capture of individuals from different sources.} We considered the case of non-attractive traps, corresponding to disjoint sets $\theta_\tau,$ $\tau=1,\ldots, J$ in the study site. We assumed that the expected number of individuals $C_\tau$ captured in a trap $\theta_\tau$ was proportional to the cumulated population in $\theta_\tau:$
\be
C_\tau=\beta_\tau\, \int_{\theta_\tau}w_\infty(x)\, dx, \hbox{ with } w_\infty(x)=\int_0^{\infty}u(t,x)\, dt,
\ee
and $\beta_\tau$ the capture rate (number of captured individuals per unit of time per unit of area in the trap $\theta_\tau$). Note that $u(t,x)\le e^{-t/\nu}\, \max u_0;$ this means that the value of $w_\infty(x)$ can be precisely approached by computing the above integral over a finite interval. For sufficiently small traps, $w_\infty(x)$ can be considered constant in $\theta_\tau,$ which leads to:
\be
C_\tau=\beta_\tau\, |\theta_\tau|  \,  w_\infty(x_\tau),
\label{eq:C_tau}
\ee
where $x_\tau$ is the location of the center of the trap $\theta_\tau$, and $ |\theta_\tau| $ is the trap area.

\begin{remark} With this approach, the trapping process has no influence on the species dynamics, i.e., the trapped individuals are not removed from the system. To avoid this lack of realism, a sink term $- \beta_\tau u \mathds{1}_{x\in \theta_\tau}$ could be added to the right-hand side of \eqref{eq:RDu}. For the sake of simplicity, we assumed that the traps were small enough to consider that this term could be neglected.
\end{remark}

Consider now the density $u^{h}(t,x)$ of individuals coming from a given habitat $\Omega^h$.  Since all the individuals are supposed to share the same dispersal and death characteristics independently of their origin,  the densities $u^{h}(t,x)$ satisfy \citep{RoqGar12}:
\be \label{eq:RDuc}
\frac{\partial u^h}{\partial t}=\Delta (D(x) \, u^h) -\frac{u^h}{\nu}, \ t>0, \, x\in \Omega,
\ee
and
\be \label{eq:predis_h}
u^h(0,x)=u_0^h(x)=\alpha \, \mathds{1}_{x\in \Omega^h}, \hbox{ for all }x\in \Omega,
\ee
with the same boundary conditions as the total population $u.$ The dynamics of the different fractions $u^{h}$ of the total population $u$ is therefore described by a system of $H$ decoupled reaction-diffusion equations. Summing up all these equations, it can be checked that for all $t,$ $x$, $$u(t,x)=\sum\limits_{h=1}^{H}u^{h}(t,x).$$The expected number of individuals coming from a habitat $\Omega^h$ and which are captured in a trap $\theta_\tau$ is then given by:
\be \label{eq:C_tauc}
C_\tau^h=\beta_\tau\, |\theta_\tau| \, w_\infty^h(x_\tau), \hbox{ with }w_\infty^h(x)=\int_0^{\infty}u^h(t,x)\, dt.
\ee
We assumed that the population was large enough so that the number of captured individuals was larger than a constant $G$ in all traps. This constant corresponds to the number of individuals genotyped in each trap.

\section{Parameters and data \label{sec:param}}

Our goal was to estimate the diffusion parameters $D(x)$ for all $x\in \Omega$. Other unknown parameters are the pre-dispersal density in the habitats ($\alpha$) and the capture rates in the different traps ($\beta_\tau$).

We assumed that the global population before dispersal, $u_0,$ was organised into several subpopulations each of which was at Hardy-Weinberg equilibrium and linkage equilibrium among loci. For the sake of clarity, we assumed that there were exactly $H$ subpopulations, with densities $u_0^h= \alpha \,\mathds{1}_{x\in \Omega^h}$, each one corresponding to a habitat $\Omega^h.$ More complex assumptions are also possible, see Remark~\ref{rem:q_vs_c}.

The positions of the habitats $\Omega^h$ and of the traps $\theta_\tau$ were known. For each subpopulation $h$ and each  locus $\lambda$ (e.g. microsatellites) out of $\Lambda$ loci,  the pre-dispersal frequencies of $A_\lambda$ alleles were known and designated as:
    \be
    \mathcal{F}_{h\lambda}=(p_{h\lambda a})_{a=1,\ldots,A_\lambda}.
    \ee
The individuals captured in $\theta_\tau$ were genotyped at the same $\Lambda$ loci. These individuals were assumed to be diploid; thus, each genotype was described by:
$$\mathcal{G}=\left\{(a^1_{\lambda},a^2_{\lambda})\right\}_{\lambda=1,\ldots,\Lambda}.$$

\section{Computation of the likelihood \label{sec:like}}

The computation of genotype likelihoods builds on a combination of classical genetic assignment studies \citep{PaeCal95,PriSte00} and seed dispersal analyses from trap data \citep{RobGar07,KleBon13}.

Among the individuals captured in a trap $\theta_\tau$, $G$ individuals have been genotyped. This led to $G$ genotypes $\mathcal{G}_{i\tau}$, $i=1,\ldots,G.$

The conditional probability that an individual $i$ carries alleles $(a^1,a^2)\in \{1,\ldots,A_\lambda\}^2$  at locus $\lambda,$ given that this individual comes from a habitat $\Omega^h$, can be deduced from the allele frequencies in subpopulation $h$ (see Section~\ref{sec:param}). The two alleles being independent, which follows from the Hardy-Weinberg equilibrium assumption in the subpopulation $u_0^h$, we get:
\be
\mathds{P}((a^1,a^2)|\Omega^h)=2^{k_\lambda} \,  p_{h\lambda a^1} \, p_{h\lambda a^2},
\ee
where $k_\lambda=0$ if the individual is homozygous at locus $\lambda$ ($a^1=a^2$) and $k_\lambda=1$ otherwise. Using the linkage equilibrium assumption among loci, we get the conditional probability of genotype $\mathcal{G}_{i\tau}$:
\be \label{eq:P_G_Omegac}
\mathds{P}(\mathcal{G}_{i\tau}|\Omega^h)=2^{k_i}\prod\limits_{\lambda=1}^{\Lambda}p_{h \lambda a^1} \, p_{h \lambda a^2},
\ee
where $k_i$ is the number of heterozygous loci in the genotype $\mathcal{G}_{i\tau}.$

\begin{remark} For the sake of simplicity, the dependence of $a^1$ and $a^2$ with respect to the locus $\lambda$ and the individual $i$ have been dropped in our notations. For instance, in formula \eqref{eq:P_G_Omegac}, $a^1$ and $a^2$ may designate different alleles, depending on the locus $\lambda$ and on the individual $i.$
\end{remark}

The law of total probability leads to:
\be \label{eq:totalproba}
\mathds{P}(\mathcal{G}_{i\tau})=2^{k_i}\sum\limits_{h=1}^{H}\left[ \prod\limits_{\lambda=1}^{\Lambda}p_{h \lambda a^1} \, p_{h \lambda a^2}\right]\mathds{P}(\hbox{indiv. }i \hbox{ comes from }\Omega^h).
\ee

We have seen in Section~\ref{sec:RD} that the expected number of individuals trapped in $\theta_\tau$ was given by $C_\tau$, and  the expected number of individuals coming from a habitat $\Omega^h$ and which are captured in a trap $\theta_\tau$ was given by $C^h_\tau.$ Let us denote by $I^h_\tau$ the number of individuals coming from $\Omega^h$ and captured in $\theta_\tau$, and let us set \be
I_\tau = \sum\limits_{h=1}^{S} I^h_\tau \ge G,
\ee
the total number of individuals captured in $\theta_\tau.$ Assume that $I^h_\tau$ follows a Poisson distribution:
\be \label{eq:poissonIctau}
I^h_\tau\sim \mathcal{P}(C^h_\tau).
\ee
Thus, $I_\tau$ also follows a Poisson distribution with parameter $$C_\tau=\sum\limits_{h=1}^{S} C^h_\tau.$$It can be verified that the conditional distribution of $I^h_\tau$ given $I_\tau$ satisfies a binomial distribution with parameters $I_\tau$ and $C^h_\tau/C_\tau.$ Thus, the conditional expectation of the proportion $I^h_\tau/I_\tau$ given $I_\tau$ is $C^h_\tau/C_\tau,$ which is independent of $I_\tau.$ Finally, this shows that:
$$\mathds{E}\left(\frac{I^h_\tau}{I_\tau}\right)=\frac{C^h_\tau}{C_\tau}.$$

The genotyping process corresponds to a sampling without replacement, this means that the number of genotyped individuals coming from habitat $\Omega^h$ follows a multivariate hypergeometric distribution with parameters $I_\tau,$ $I^h_\tau$ and $G$. For large values of $I_\tau$, this distribution converges to the multinomial distribution with parameters $G$ and $(I^1_\tau/I_\tau,\ldots, I^H_\tau/I_\tau)$.  Using this multinomial distribution, we can compute the probability that a genotyped individual $i$ trapped in $\theta_\tau$ comes from a habitat $h$:
\be \label{eq:Pindivi_omega_h}
\mathds{P}(\hbox{indiv. }i \hbox{ comes from }\Omega^h)=\mathds{E}\left(\frac{I^h_\tau}{I_\tau}\right)=\frac{C^h_\tau}{C_\tau}.
\ee
The hypergeometric distribution would lead to the same formula, but the advantage of the multinomial distribution is that it guarantees an independence assumption between the individuals trapped at a same location.

Assuming that the trapping and genotyping processes are independent, and using formulas \eqref{eq:totalproba} and \eqref{eq:Pindivi_omega_h}, we finally get the likelihood function associated with the unknown parameter $D$, given the genotypes $\mathcal{G}_{i\tau}$:
\be \baa{ll}
\mathcal{L}(D) & =  \prod\limits_{\tau=1, \ldots,  J} \prod\limits_{i=1, \ldots, G}\mathds{P}(\mathcal{G}_{i\tau}| D) \\
& = \prod\limits_{\tau=1, \ldots,  J} \prod\limits_{i=1, \ldots, G} \mathds{P}(\mathcal{G}_{i\tau}| C^h_\tau/C_\tau ) \\
& =2^{k}  \prod\limits_{\tau=1, \ldots,  J} \prod\limits_{i=1, \ldots, G} \sum\limits_{h=1}^{H}\left[ \frac{C^h_\tau}{C_\tau}\prod\limits_{\lambda=1}^{\Lambda}p_{h \lambda a^1} \, p_{h \lambda a^2}\right],
\eaa
\label{eq:like}
\ee
where $k$ is the total number of heterozygous loci in the genotyped population. Coming back to the definitions \eqref{eq:C_tau} and \eqref{eq:C_tauc} of $C_\tau$ and $C^h_\tau,$ we can compute the ratio:
\be \label{eq:ratioC_tau}
\ds \frac{C^h_\tau}{C_\tau} = \frac{w_\infty^h(x_\tau)}{w_\infty(x_\tau)}.
\ee
This shows that $C^h_\tau/C_\tau$ is independent of the capture rates $\beta_\tau.$ From the linearity of the equations \eqref{eq:RDu} and \eqref{eq:RDuc}, it follows that $w_\infty^h$ and $w_\infty$ are proportional to $\alpha$ which means that the ratio $C^h_\tau/C_\tau$ is also independent of the choice of $\alpha$. Thus, the likelihood $\mathcal{L}(D)$ can be computed with an arbitrary choice of parameters,  e.g.  $\alpha,\beta_\tau=1$. Note that if the source intensity $\alpha$ was spatially variable, it would not simplify in the expression $C^h_\tau/C_\tau.$ In such case, $\alpha$ should be estimated in each source, which is only possible up to a multiplicative constant since $C^h_\tau$ would be proportional to the value of $\alpha$ in each habitat $h$.


\begin{remark}\label{rem:q_vs_c}
We recall that $u_0$ was organised into several subpopulations at Hardy-Weinberg equilibrium. Here, we assumed that, before dispersal, these subpopulations coincided with the habitats, which were disjoint. A first easy generalization would be to consider several subpopulations in the same habitat. Assume that there are $R$  subpopulations with pre-dispersal densities $u_0^r(x)=\sum\limits_{h=1}^H \mu^{h r}\,\mathds{1}_{x\in \Omega^h},$ and that the allele frequencies $\mathcal{F}_{r\lambda}$ are known in these subpopulations.
In this case, the probability $\mathds{P}(\mathcal{G}_{i\tau}|\Omega^h)$ does not satisfy formula~\eqref{eq:P_G_Omegac} but can be computed as follows:
\be
\mathds{P}(\mathcal{G}_{i\tau}|\Omega^h)=\sum\limits_{r=1}^R  \mu^{h r}\, \mathds{P}(\mathcal{G}_{i\tau}|\hbox{indiv. }i \hbox{ comes from subpop. }r),
\ee
which leads to:
\be
\mathds{P}(\mathcal{G}_{i\tau}|\Omega^h)=2^{k_i}\sum\limits_{r=1}^R \mu^{h r}\, \prod\limits_{\lambda=1}^{\Lambda}p_{r \lambda a^1} \, p_{r \lambda a^2}.
\ee
\end{remark}

\section{Numerical computations \label{sec:num}}

The aim of this section was to validate the maximum likelihood estimator of Section~\ref{sec:like} on a simulated data set.

\subsection{Simulated data set \label{sec:simu_data}}

\begin{figure}
\centering
\includegraphics[width=0.5\textwidth]{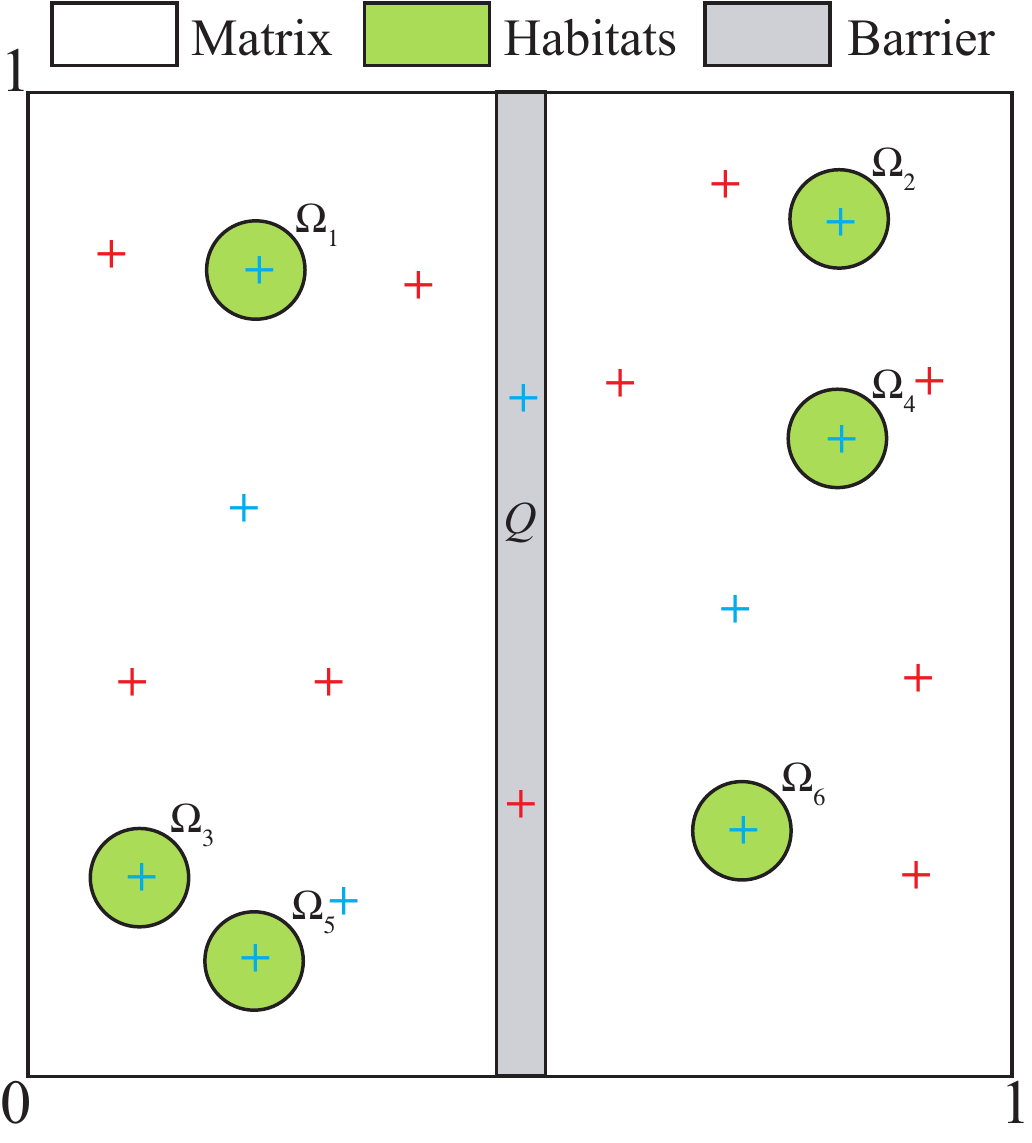}
\caption{Study site $\Omega=(0,1)\times (0,1)$ with six habitats $\Omega^h$ of centers $x_h$, for $h=1,\ldots,6$. The grey region $Q$ corresponds to a barrier to dispersal. Blue crosses correspond to the positions of the first 10 traps $\theta_\tau,$ $\tau=1,\ldots,10$; red crosses correspond to the positions of a supplementary set of ten additional traps $\theta_\tau,$ $\tau=11,\ldots,20.$ }
\label{fig:land}
\end{figure}

\noindent{\it Landscape.} The study site $\Omega$ was a unit square $(0,1)\times (0,1)$ containing $H=6$ habitats $\Omega^h$, described by balls $B_R(x_h)$ of radius $R=0.05$ and centers $x_h \in \Omega$ and a rectangular region $Q=(q-R/2,q+R/2)\times (0,1)$ modelling a barrier to dispersal ($q=0.5$). The rest of the study site was considered as the matrix (Fig~\ref{fig:land}).

The general idea was to consider diffusion parameters $D_1$ in the matrix, $D_2=D_1/2$ in the habitats and $D_3=D_1/10$ in the barrier. However, for the well-posedness of the reaction-diffusion equations for $u$ and $u^h$, the coefficient $D(x)$ had to be positive and smooth. We thus defined the heterogeneous diffusion parameters $D(x)$ as:
\be D(x)=\exp\lp d_1 + d_2 \, \sum\limits_{h=1}^H \phi(x-x_h)+ d_3 \, \psi(x)\rp,\label{eq:diff}\ee for smooth positive functions $\phi$ and $\psi$,  such that $\phi(x-x_h)$ was compactly supported in $B_{2\, R}(x_h)$ for any $h=1,\ldots, S$ and $\psi$ was compactly supported in $(q-R,q+R)\times (0,1)$ and $\max \phi=\max \psi=1$. The precise shape of $\phi$ and $\psi$ is detailed in Appendix~B. The numerical values of $d_1,$ $d_2$ and $d_3$ that we used in our computations were:
\be \label{eq:true} d_1=\log(0.01), \ d_2=-\log(2), \hbox{ and }d_3=-\log(10).\ee
With this framework, the diffusion parameters were equal to $D_1=10^{-2}$ in the matrix, far from the habitats and from the barrier, to $D_2=D_1/2$ at the center of the habitats and to $D_3=D_1/10$ at the center of the barrier. See Remark~\ref{rem:diff} for some comments on these parameter values.

\begin{remark}\label{rem:diff}
Assuming that the unit square corresponds to a 10km $\times$ 10km region, and that the unit of time is one day, a diffusion parameter $D=10^{-2}$ corresponds to $1$km$^2/$day. Using the formula \citep[see e.g.][]{Tur98,Roq13}:$$D= \frac{(\hbox{length of a straigth line move during one time step})^2}{4\times \hbox{duration of the time step}},$$for random walk movements with one direction change per minute this value of the diffusion parameter corresponds to a flying speed of 53m per minute. Under the same assumptions, $D=10^{-2}/2$ and $D=10^{-2}/10$ correspond to flying speeds of $37$m and $17$m per minute, respectively. See \cite{Kar83} for reference values of diffusion parameters of several insect species.
\end{remark}

\

\noindent {\it Allele frequencies.} The number of loci and the number of alleles at each locus were $\Lambda=10$ and $A=10$. Following \cite{PriSte00}, we drew the allele frequencies in a flat Dirichlet distribution, with concentration parameter $q$ in each subpopulation $h$ and for each locus $\lambda:$
\be
\mathcal{F}_{s\lambda}\sim \mathcal{D}(q).
\ee
Values of $q$ larger than 1 lead to evenly distributed frequencies, whereas smaller values of $q$ lead to distributions which are concentrated on a few components. In our simulations, we adjusted the value of $q$ such that the fixation index $F_{ST}$ among subpopulations (see Appendix~C) was close to 0.01,  0.05 or 0.1 (Table~\ref{tab:FST}), with a relative tolerance of $0.1\%$.


\

\begin{figure}
\centering
\subfigure[$h=1$]{\includegraphics[width=0.3\textwidth]{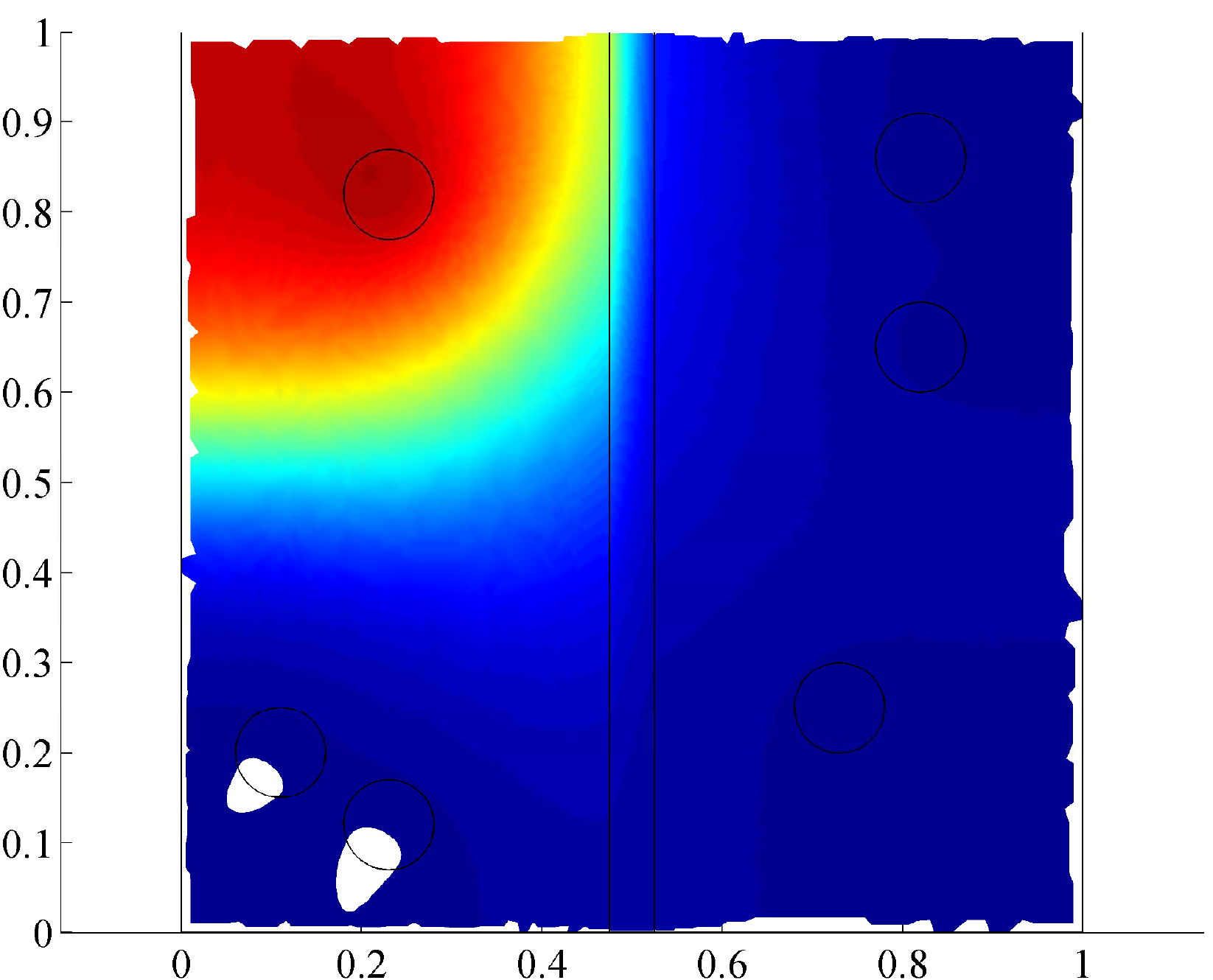}}
\subfigure[$h=2$]{\includegraphics[width=0.3\textwidth]{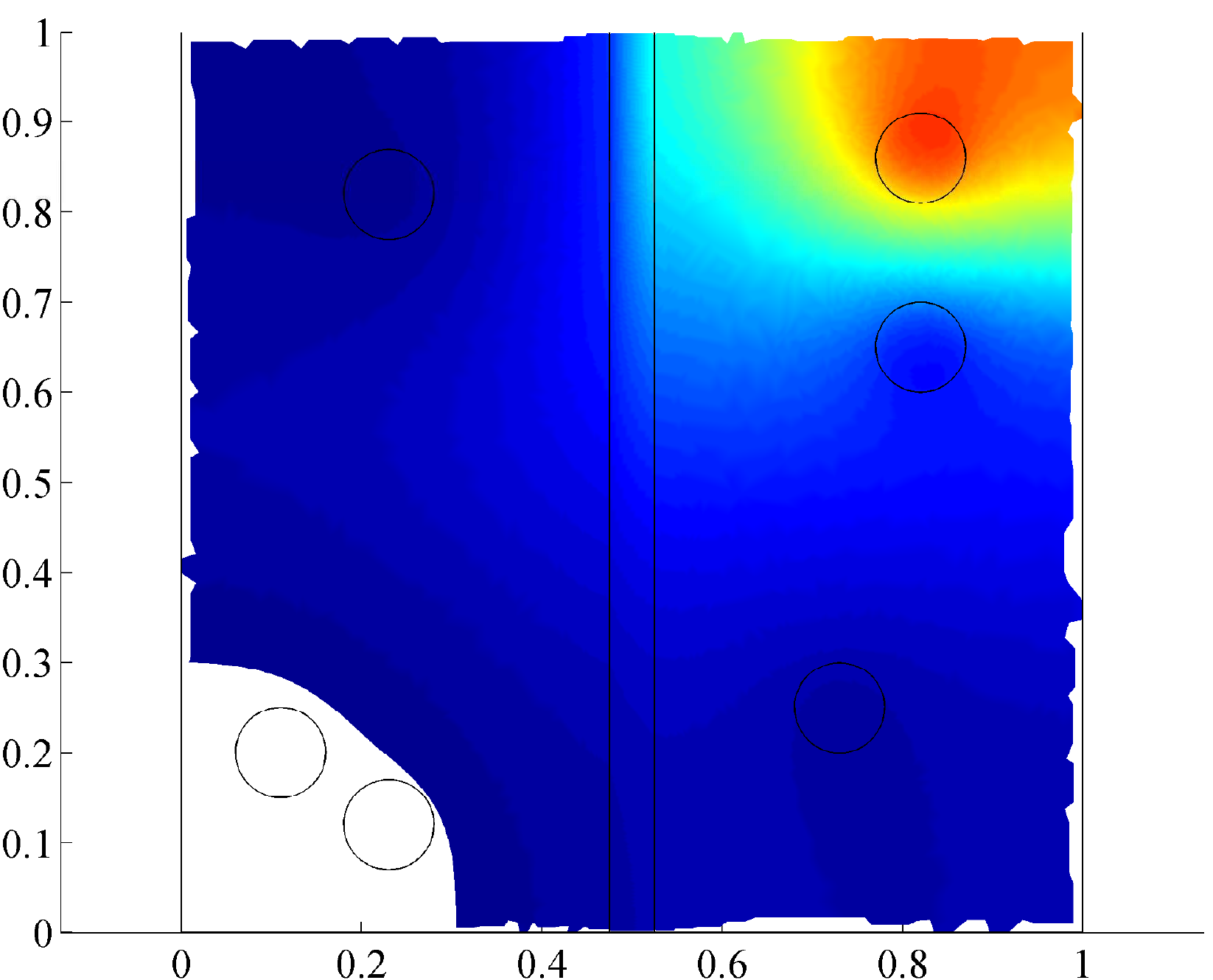}}
\subfigure[$h=3$]{\includegraphics[width=0.3\textwidth]{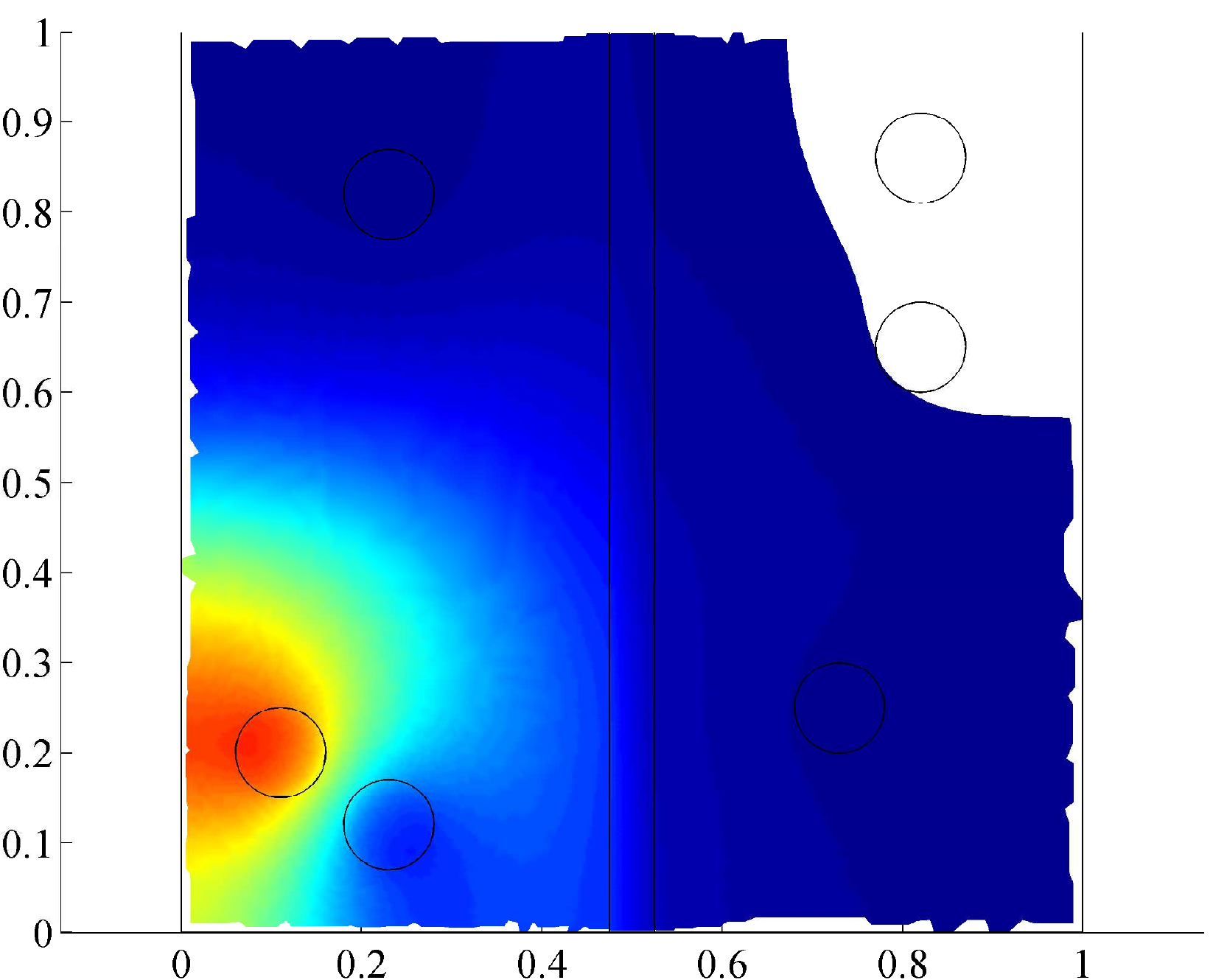}}
\subfigure[$h=4$]{\includegraphics[width=0.3\textwidth]{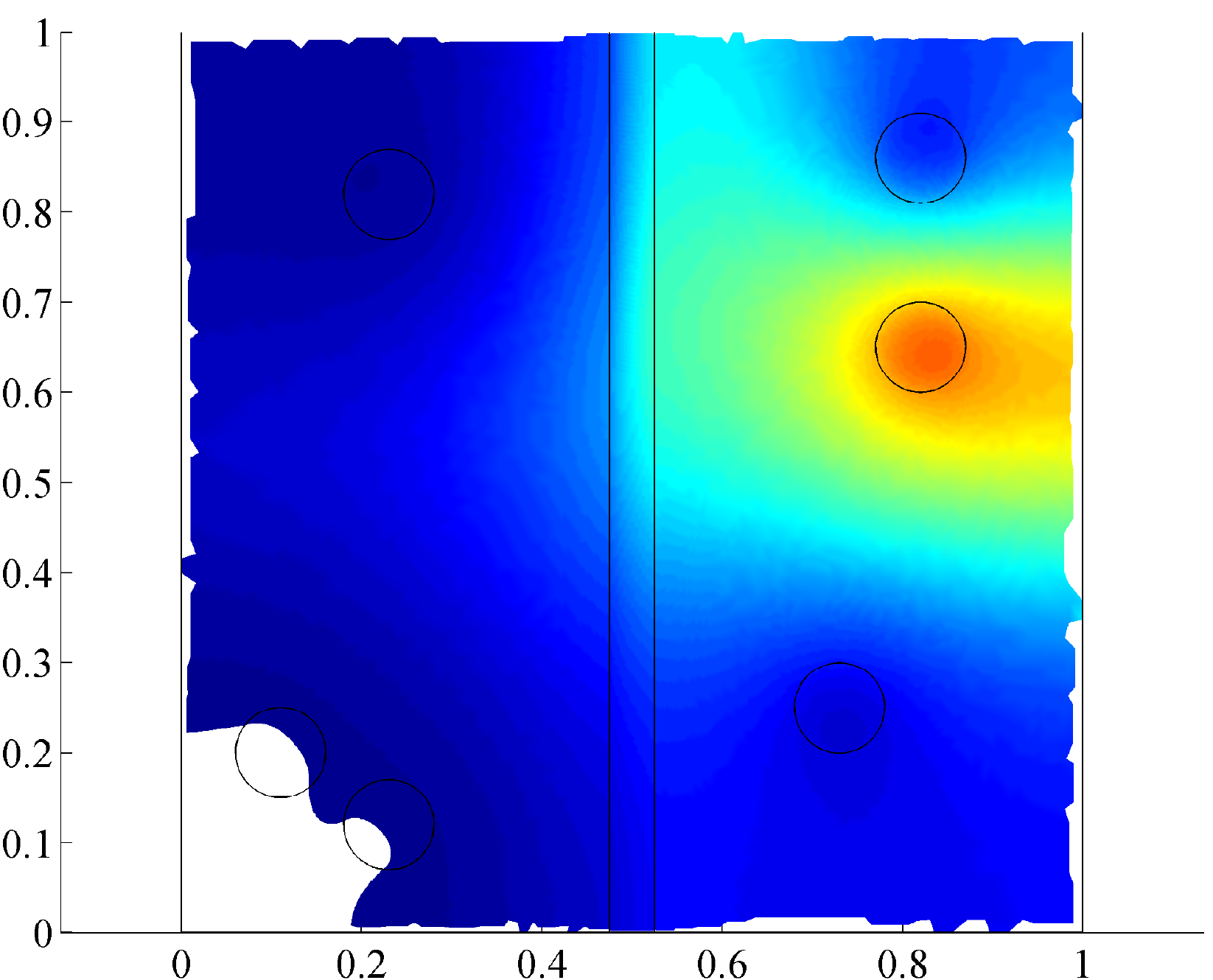}}
\subfigure[$h=5$]{\includegraphics[width=0.3\textwidth]{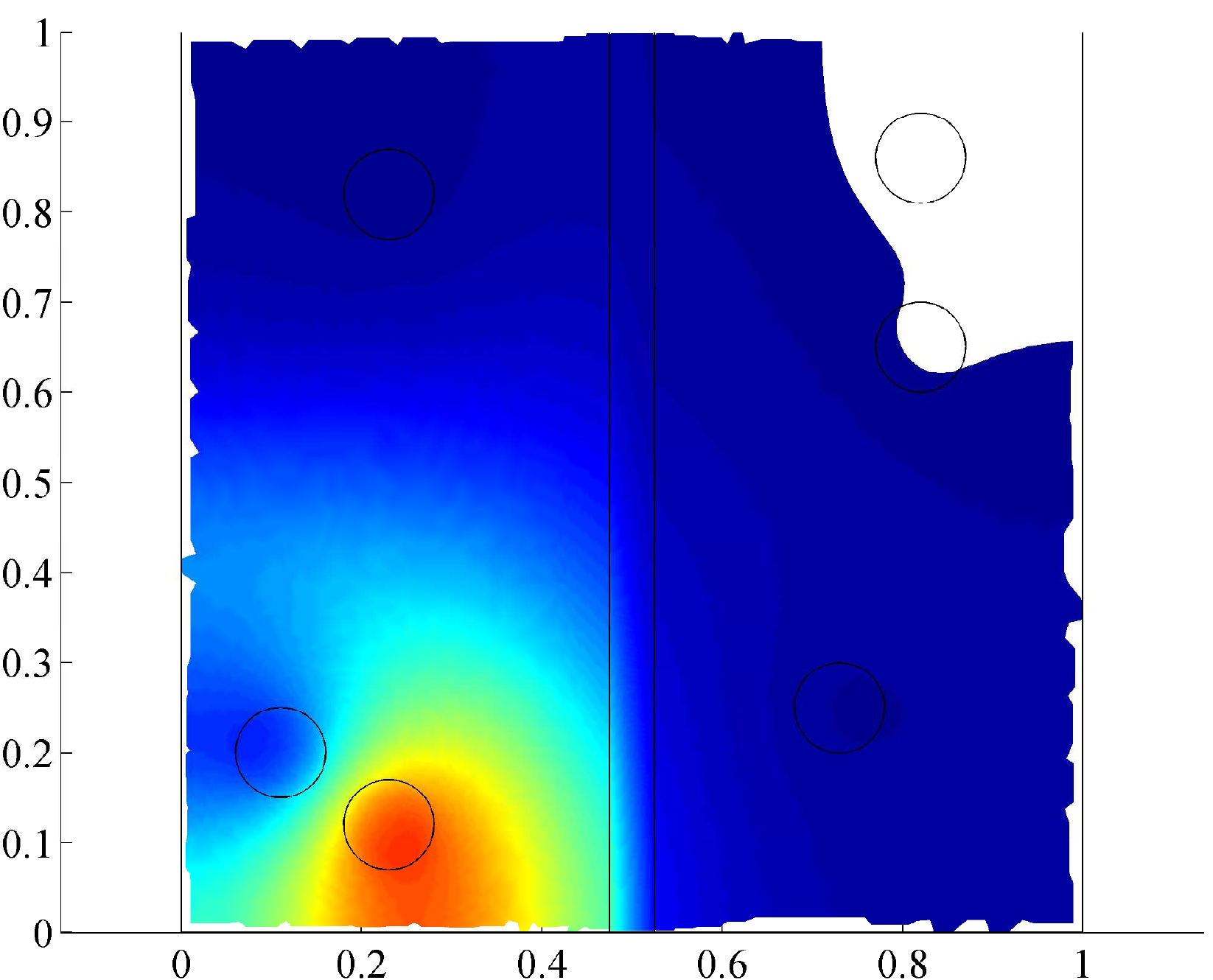}}
\subfigure[$h=6$]{\includegraphics[width=0.3\textwidth]{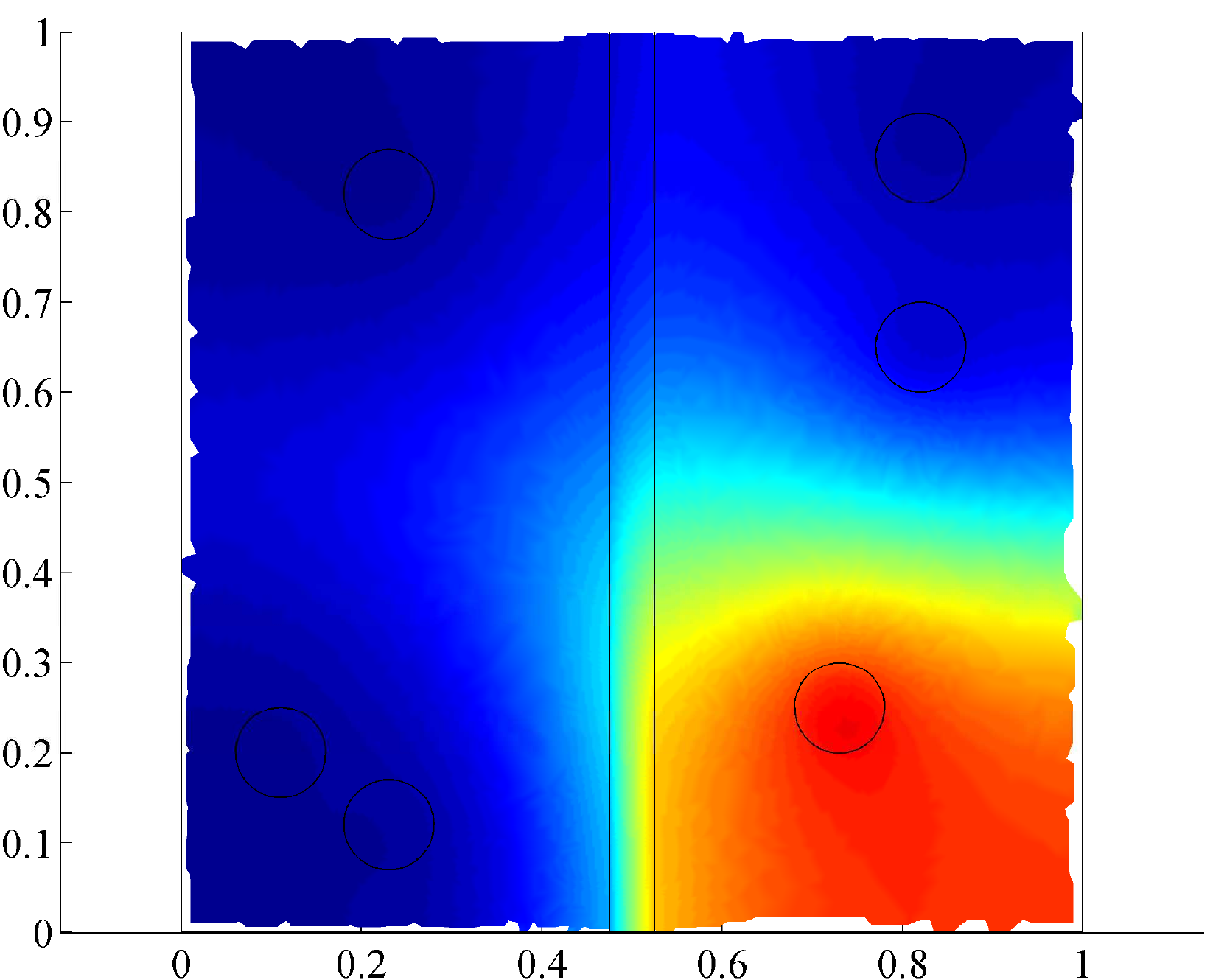}}
\includegraphics[width=0.035\textwidth]{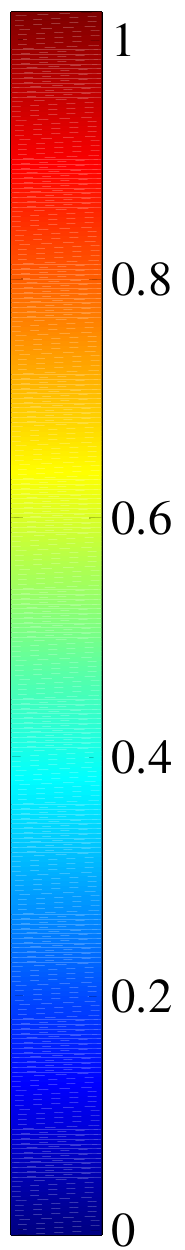}
\caption{Probability $w_\infty^h(x)/w_\infty(x)$ that an individual trapped at the position $x\in \Omega$ comes from the habitat $\Omega^h$; see formula \eqref{eq:ratioC_tau}. White regions indicate that the probability is smaller than $0.005$. The value of $w_\infty^h(x)/w_\infty(x)$ has been computed at $t=20,$ with the true parameter values (see Appendix~D).}
\label{fig:P}
\end{figure}

\noindent {\it Simulation of the measured data.} We solved the reaction-diffusion models for the cumulated population densities  $w_\infty$ and $w_\infty^h$, for $h=1,\ldots,6$ with the true diffusion parameter values \eqref{eq:true} and $\nu=5$ for the life expectancy of the dispersers (see Appendix~D). The proportions $(C_\tau^h/C_\tau)_{h=1,\ldots,6}$ have been computed using formulas  \eqref{eq:C_tau} and \eqref{eq:C_tauc}. We recall that, from formulas \eqref{eq:Pindivi_omega_h} and \eqref{eq:ratioC_tau}, $C_\tau^h/C_\tau=w_\infty^h(x_\tau)/w_\infty(x_\tau)$ is the probability that an individual trapped in $\theta_\tau$ comes from the habitat $\Omega^h$. The probability $w_\infty^h(x)/w_\infty(x)$ can be computed at each point in $\Omega$; it is depicted in Fig.~\ref{fig:P}. Movies of the dynamics of the probability $w_t^h(x)/w_t(x)$ that an individual trapped in $\theta_\tau$ between the times $0$ and $t$ comes from the habitat $\Omega^h$ are available as supplementary materials.

In each trap $\theta_\tau$ (see Fig.~\ref{fig:land} for the locations of the traps), the numbers of genotyped individuals coming from the habitats $\Omega^h$ followed a multinomial distribution with parameters $G$ and $(C_\tau^h/C_\tau)_{h=1,\ldots,6}$. We tested the effect of the number of traps by using $J=10$ or $20$ traps and of the total number of genotyped individuals by using $J\times G=500$ or $2000$ genotyped individuals (Table~\ref{tab:FST}).


For each genotyped individual $i$, the genotype $\mathcal{G}_{i\tau}$ was randomly drawn according to the allele frequencies $\mathcal{F}_{h\lambda}$ in its habitat of origin.
The simulated observations consisted in the genotypes $\mathcal{G}_{i\tau}$, i.e., the set:
\be
\mathcal{M}=\{\mathcal{G}_{i\tau}, \ i=1,\ldots,G, \ \tau=1,\ldots, J \}.
\ee


For each of the $12$ sets of parameters $(F_{ST},  J,  J \, G),$ we simulated $70$ data sets  $\mathcal{M}.$  For each data set, the estimator $(\hat d_1, \hat d_2, \hat d_3)$ of $(d_1,d_2,d_3)$ has been obtained by minimizing $-\log(\mathcal L (D))$ (see formula~\eqref{eq:like}). The minimization was performed using the Matlab$^\circledR$ constrained gradient-based minimization algorithm fmincon, with the constraint \be\label{eq:constraint} (\hat d_1, \hat d_2, \hat d_3) \in (d_1-5,d_1+5)\times (d_2-5,d_2+5)\times (d_3-5,d_3+5).\ee In the computation of $-\log(\mathcal L (D))$, the numerical evaluation of the quantities $(C_\tau^h/C_\tau)_{h=1,\ldots,6}$ was based on the finite element method as described in Appendix~D.
The average computation time for one estimation was about $45$min with a dual Core Intel$^\circledR$ processor, while the computation of the likelihood took about $5$sec.  The Bayesian method \citep{MarRob07} would be more computationally intensive, but it could also be used to compute a posterior distribution of $D(x)$ at each position $x$.

\subsection{Results \label{sec:results}}

\begin{figure}
\centering
\subfigure[$\tilde d_3-d_3=-2$]{\includegraphics[width=0.18\textwidth]{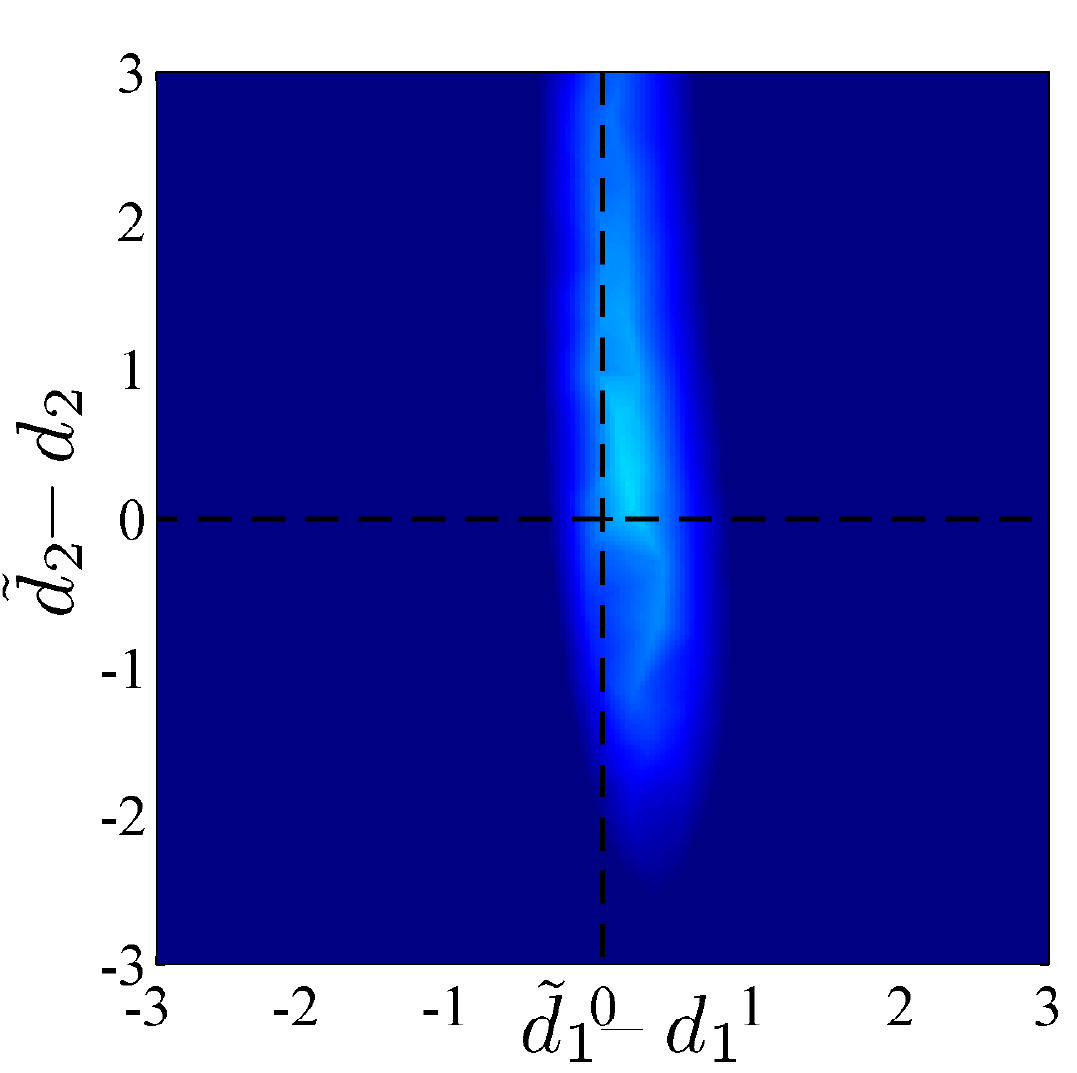}}
\subfigure[$\tilde d_3-d_3=-1$]{\includegraphics[width=0.18\textwidth]{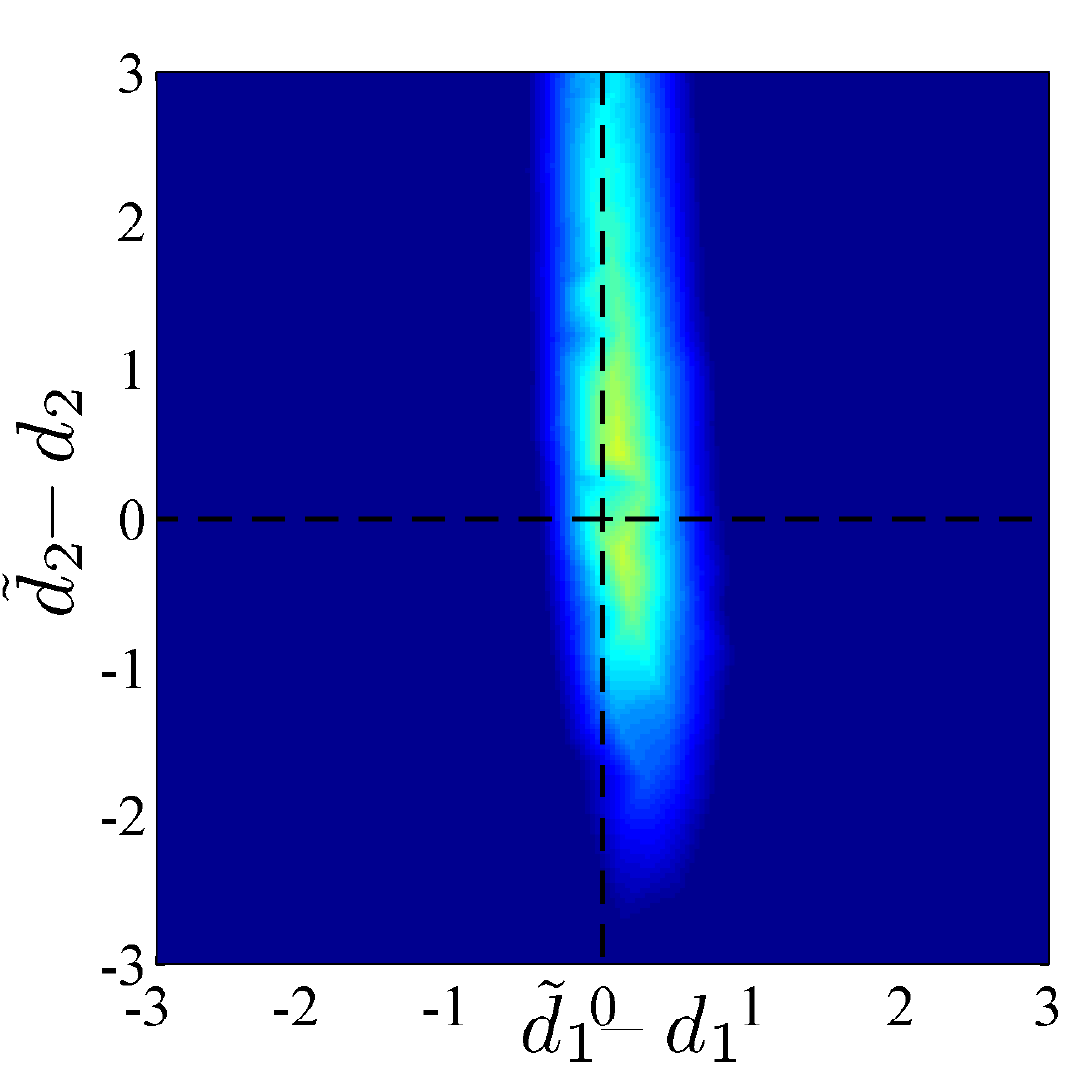}}
\subfigure[$\tilde d_3-d_3=0$]{\includegraphics[width=0.18\textwidth]{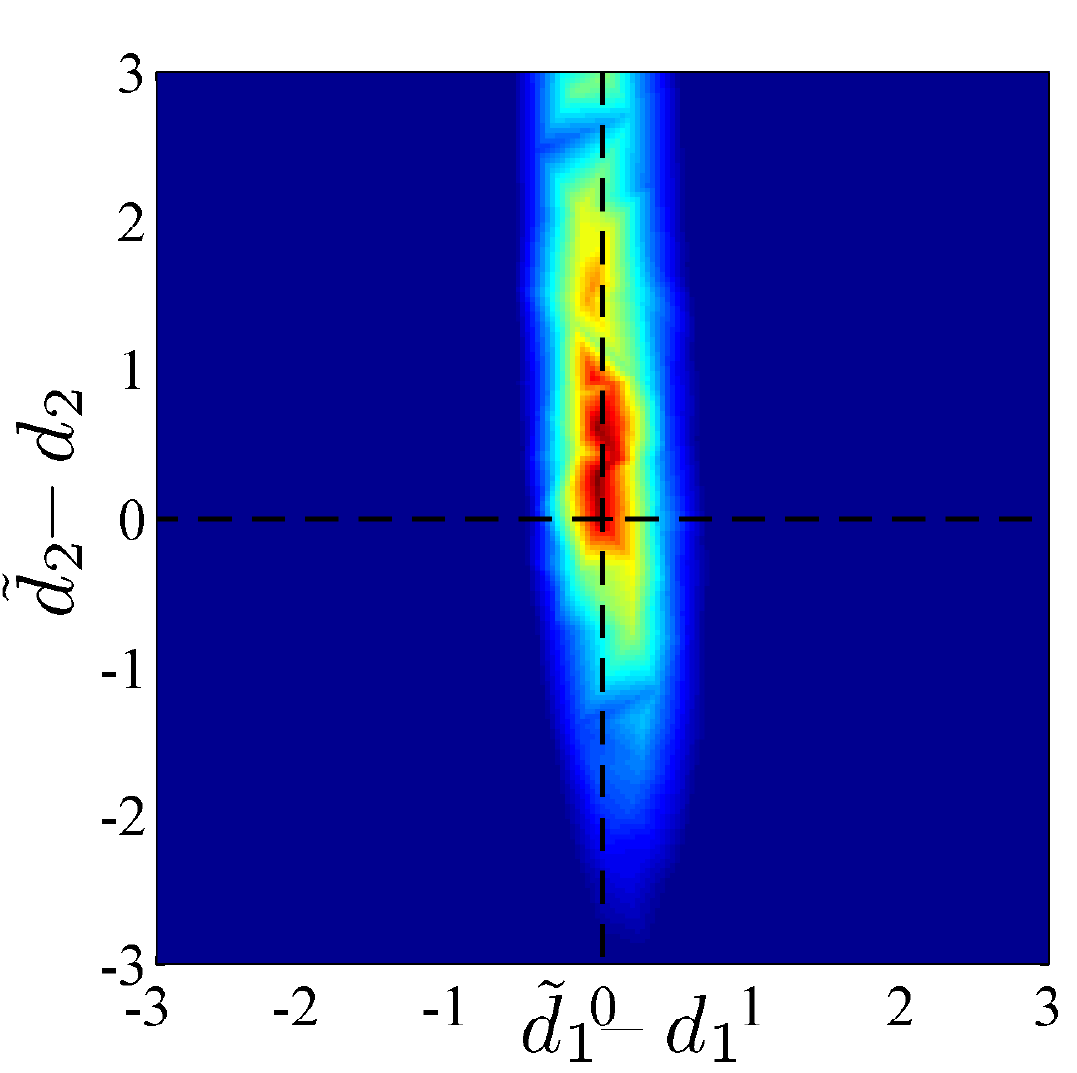}}
\subfigure[$\tilde d_3-d_3=1$]{\includegraphics[width=0.18\textwidth]{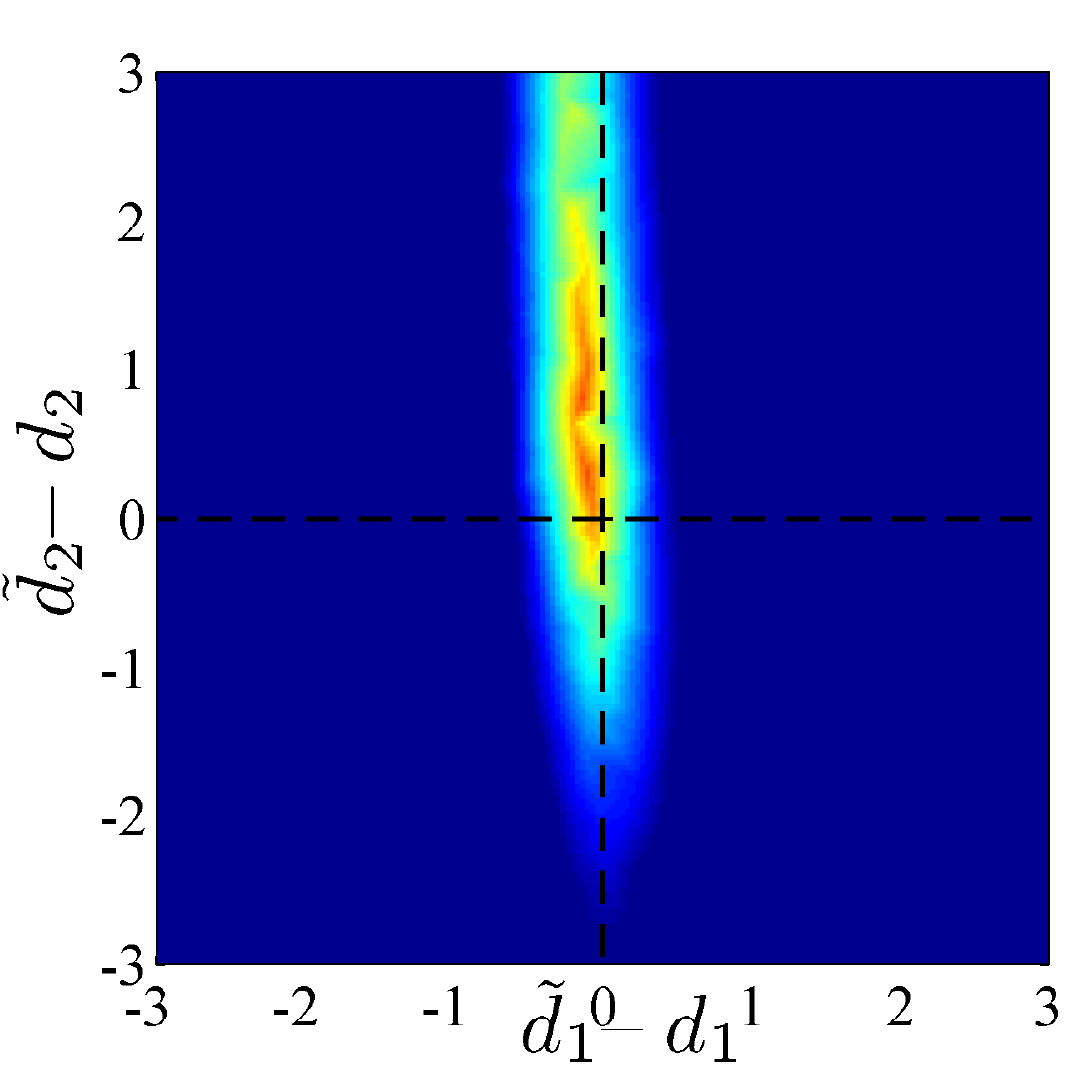}}
\subfigure[$\tilde d_3-d_3=2$]{\includegraphics[width=0.18\textwidth]{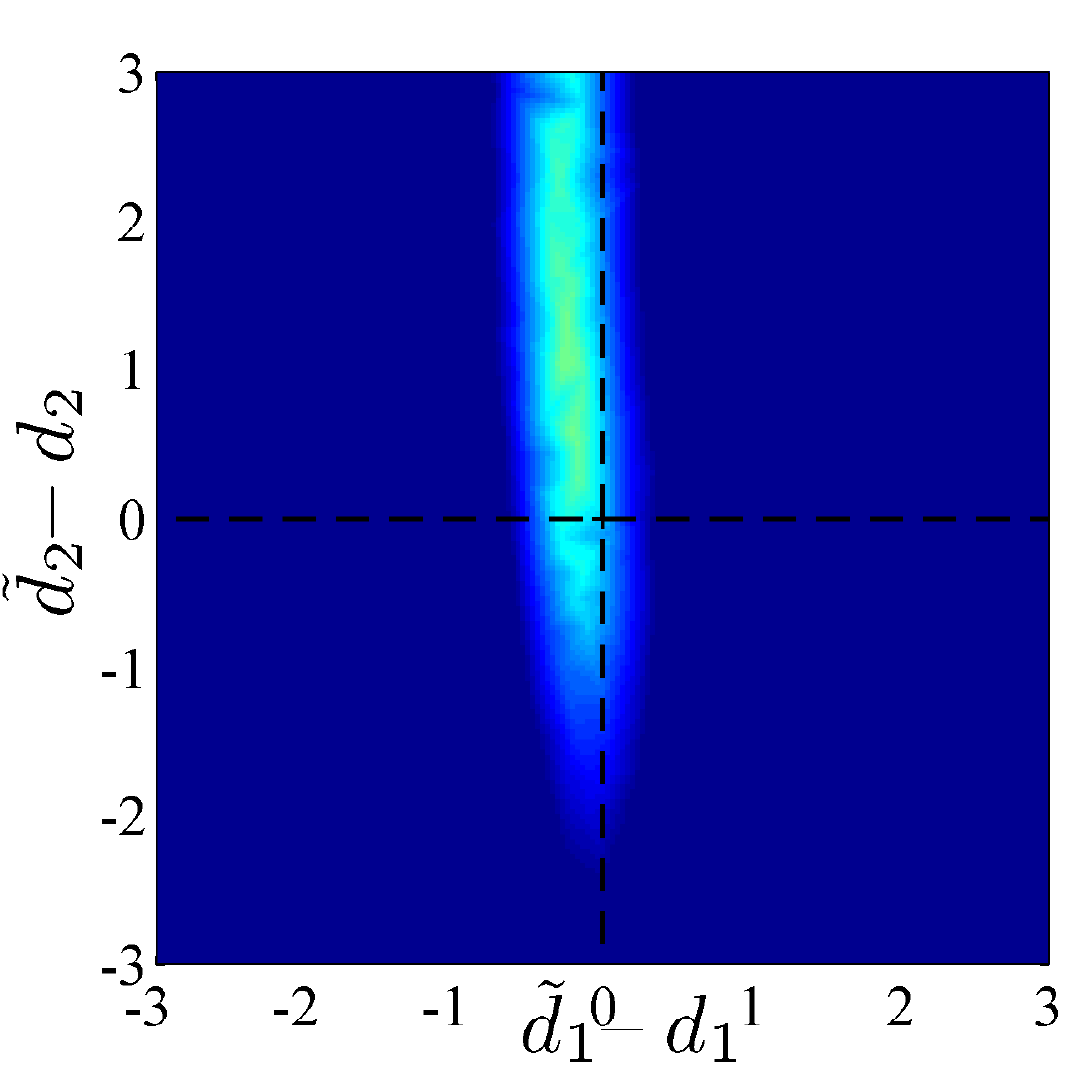}}
\includegraphics[width=0.05\textwidth]{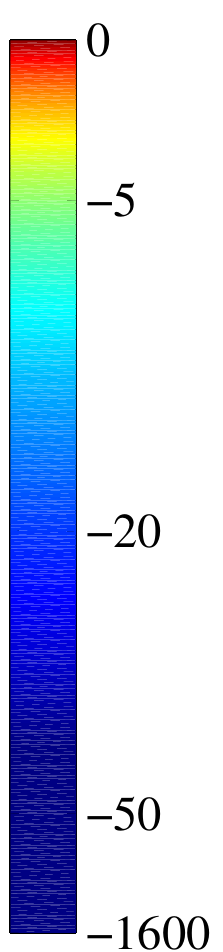}
\caption{Five sections of the 3-dimensional log-likelihood function $\ln(\mathcal{L}(\tilde d_1,\tilde d_2, \tilde d_3))$. The color scale corresponds to the gap between the log-likelihood and the maximum log-likelihood, where the maximum is taken over the parameter region $(\tilde d_1,\tilde d_2,\tilde d_3) \in (d_1-5,d_1+5)\times (d_2-5,d_2+5)\times (d_3-5,d_3+5)$.}
\label{fig:Vrais1}
\end{figure}

The typical profile of the log of the likelihood function \eqref{eq:like} is depicted in Fig~\ref{fig:Vrais1} for one simulation with $F_{ST}=0.05,$ $J=20$ traps and $G \, J=2000$ individuals. The likelihood tends to decay as $(\tilde d_1,\tilde d_2,\tilde d_3)$ diverges from the true value $(d_1,d_2,d_3),$ suggesting an efficient parameter estimation by maximum likelihood, even with the constraint \eqref{eq:constraint} was relaxed.

The direct analysis of the quality of the estimator $(\hat d_1,\hat d_2,\hat d_3)$ is presented in Table~\ref{tab:FST}, in terms of the $F_{ST}$ value, the number of traps and the number of genotyped individuals. In this section, we focus on the more biologically meaningful estimators
$$\hat D_1= \exp\lp \hat d_1\rp, \ \hat D_2=  \exp\lp \hat d_1 +\hat d_2\rp, \ \hat D_3=  \exp\lp \hat d_1 +\hat d_3\rp$$
which can be directly compared to the values $D_1$, $D_2$, $D_3$ of the diffusion parameter in each of the three regions (matrix far from the other regions, center of the habitats and center of the barrier, respectively).

\begin{figure}
\centering
\hspace{0.2cm}\includegraphics[width=0.49\textwidth]{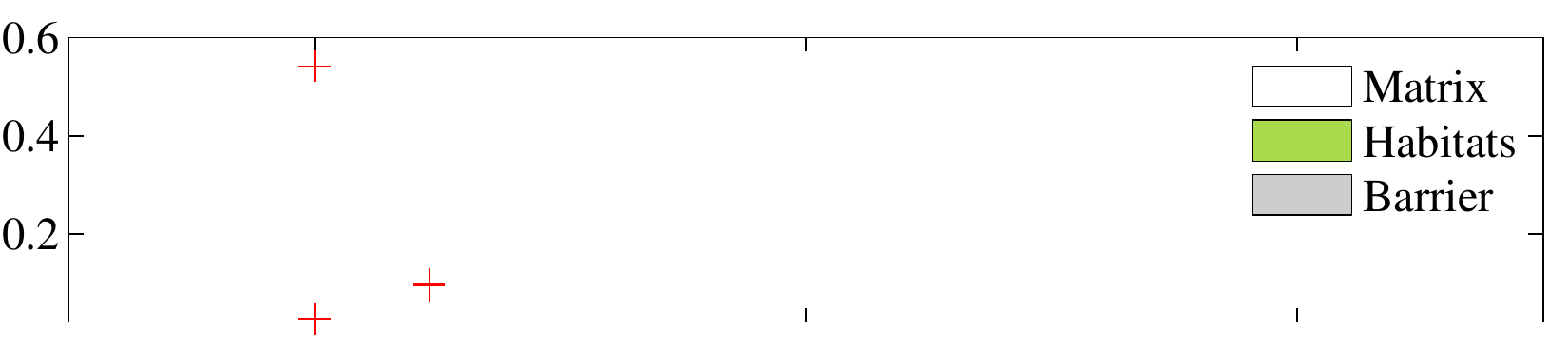}

\vspace{-0.1cm}
\includegraphics[width=0.5\textwidth]{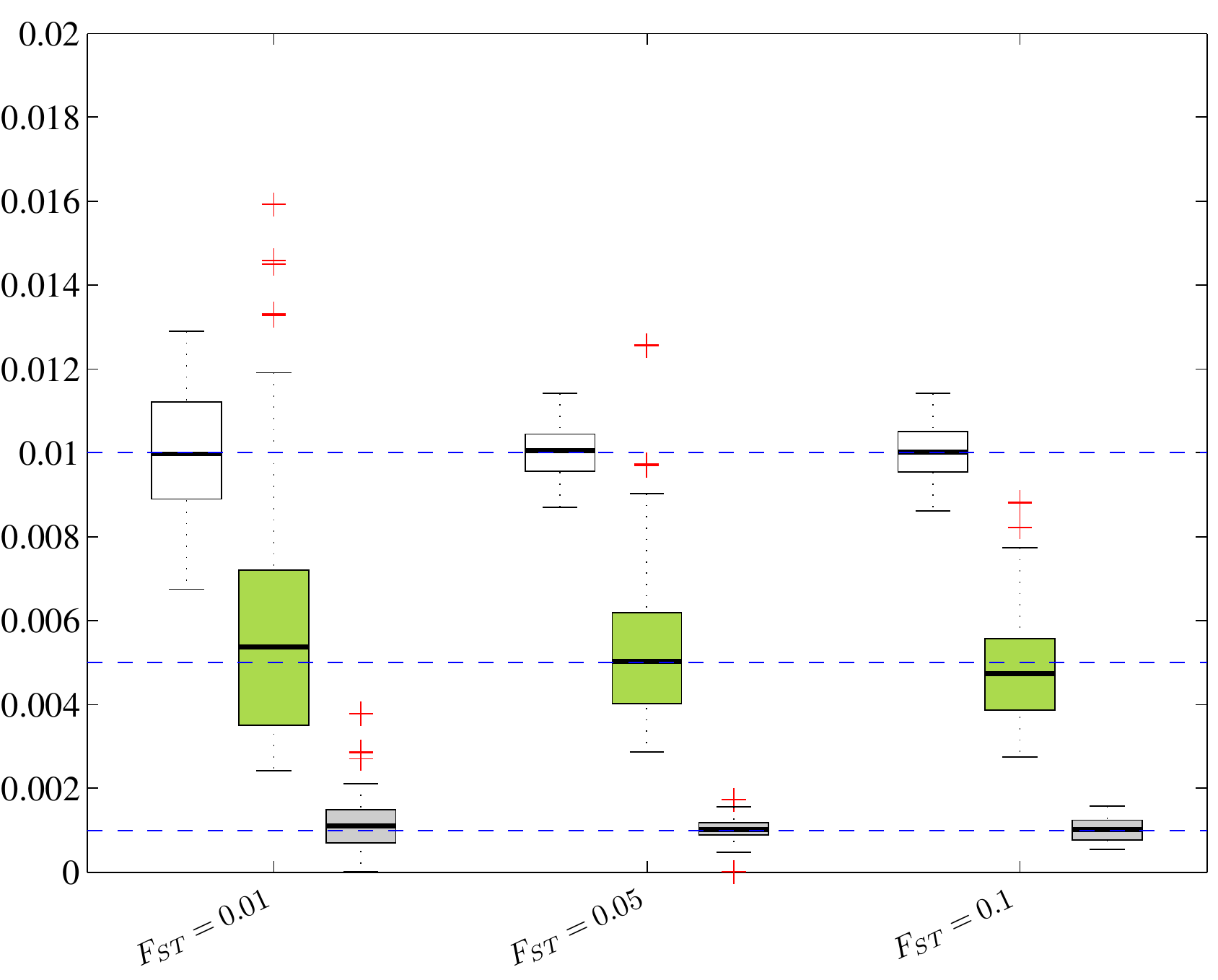}
 \caption{Effect of the $F_{ST}$ on the quality of the estimator $\hat D(x)$ in each of the three regions: $\hat D_1$ in the matrix (in white), $\hat D_2$ in the habitats (in green) and $\hat D_3$ in the barrier (in grey). The blue dashed lines correspond to the true values of $D$ in the three regions: $D_1=0.01,$  $D_2=0.005$ and $D_3=0.001.$}
 \label{fig:bp1}
\end{figure}

For each $F_{ST}$ value (0.01, 0.05 and 0.1), with $J=20$ traps and $J \, G=2000$ genotyped individuals, we observe in Fig.~\ref{fig:bp1} that the median of $\hat D$ is correctly centered on the true value of $D$ in each of the three regions (see Fig.~\ref{fig:land} and Section~\ref{sec:simu_data} for the definition of the three regions). We also note that there are no outliers among the estimated values of $D$ in the matrix (the largest region in Fig.~\ref{fig:land}) for all  $F_{ST}$ values.  Outliers far from the true value of $D$ (about 12 times the true value) appear in the other regions for small $F_{ST}$ values ($F_{ST}$=0.01). For larger $F_{ST}$ values, the number of outliers is reduced and they are closer to the true value (2 outliers in the habitats and the barrier  for $F_{ST}$=0.05 and only 2 outliers
in the habitats region for $F_{ST}$=0.1). In the three regions (matrix/habitats/barrier), the interquartile ranges are reduced as the $F_{ST}$ is increased from $0.01$ to $0.05$; this decrease of the interquartile ranges by a factor 2 in average indicates a strong effect of the $F_{ST}$ on the accuracy of the estimation. There is no clear difference  between the interquartile ranges corresponding to $F_{ST}=0.05$ and $F_{ST}=0.1;$ however, the bias and the standard deviation of $(\hat d_1,\hat d_2,\hat d_3)$ are still improved when the $F_{ST}$ is increased from $0.05$ to $F_{ST}=0.1$ (see Table~\ref{tab:FST}).

\begin{table}[h!]
   \centering
\begin{tabular}{|c|c|c|c|c|}
\hline
$F_{ST}$ &  $\sharp$traps $ J$ &  $\sharp$indiv. $ J \, G$ & Bias ($\%$ true value) & Std dev ($\%$ true value) \tabularnewline\hline

0.01 & 20 & 2000 & (-0.2, 26.2, -3.9) & (3.2, 135.4, 65.5) \tabularnewline\hline
0.05 & 20 & 2000 & (-0.1, 5.7, 0.9) & (1.7, 53.1, 33.7) \tabularnewline\hline
0.1 & 20 & 2000 & (-0.1, 0.9, 0.3) & (1.6, 49.4, 15.0) \tabularnewline\hline
0.1 & 20 & 500 &  (0.0, 15.1, -10.3) & (3.0, 114.9, 38.6) \tabularnewline\hline
0.1 & 10 & 2000 & (-0.1, 2.1, -5.7) & (2.0, 36.8, 36.0) \tabularnewline\hline
0.1 & 10 & 500 & (0.4, -10.6, -4.4) & (4.4, 75.3, 61.2) \tabularnewline\hline
 \end{tabular}
\caption{Effect of the $F_{ST},$ the number of traps and of the total number of genotyped individuals on the quality of the estimator $(\hat d_1, \hat d_2, \hat d_3)$. }
 \label{tab:FST}
 \end{table}

\begin{figure}
\centering
\includegraphics[width=0.4\textwidth]{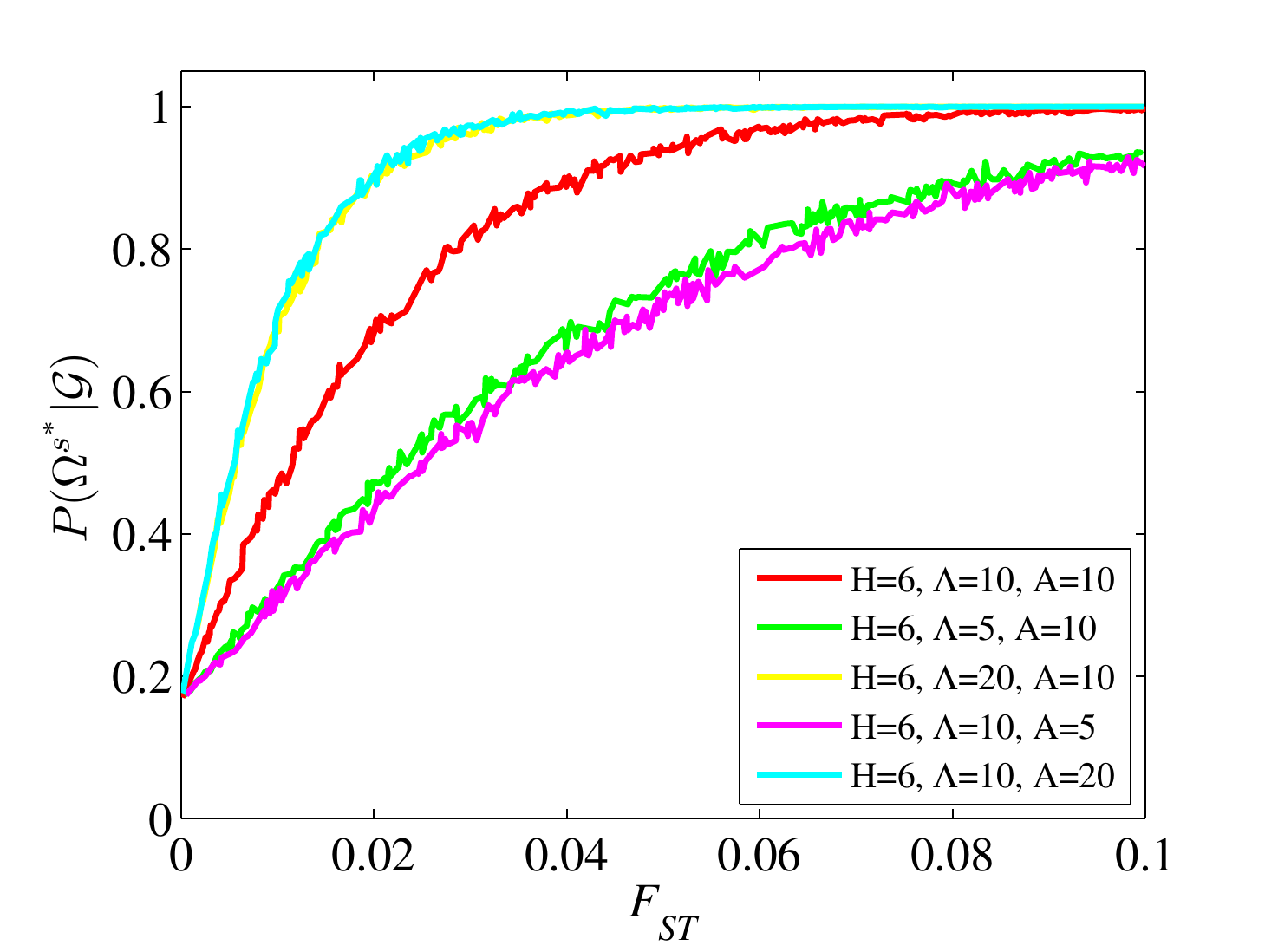}
\includegraphics[width=0.4\textwidth]{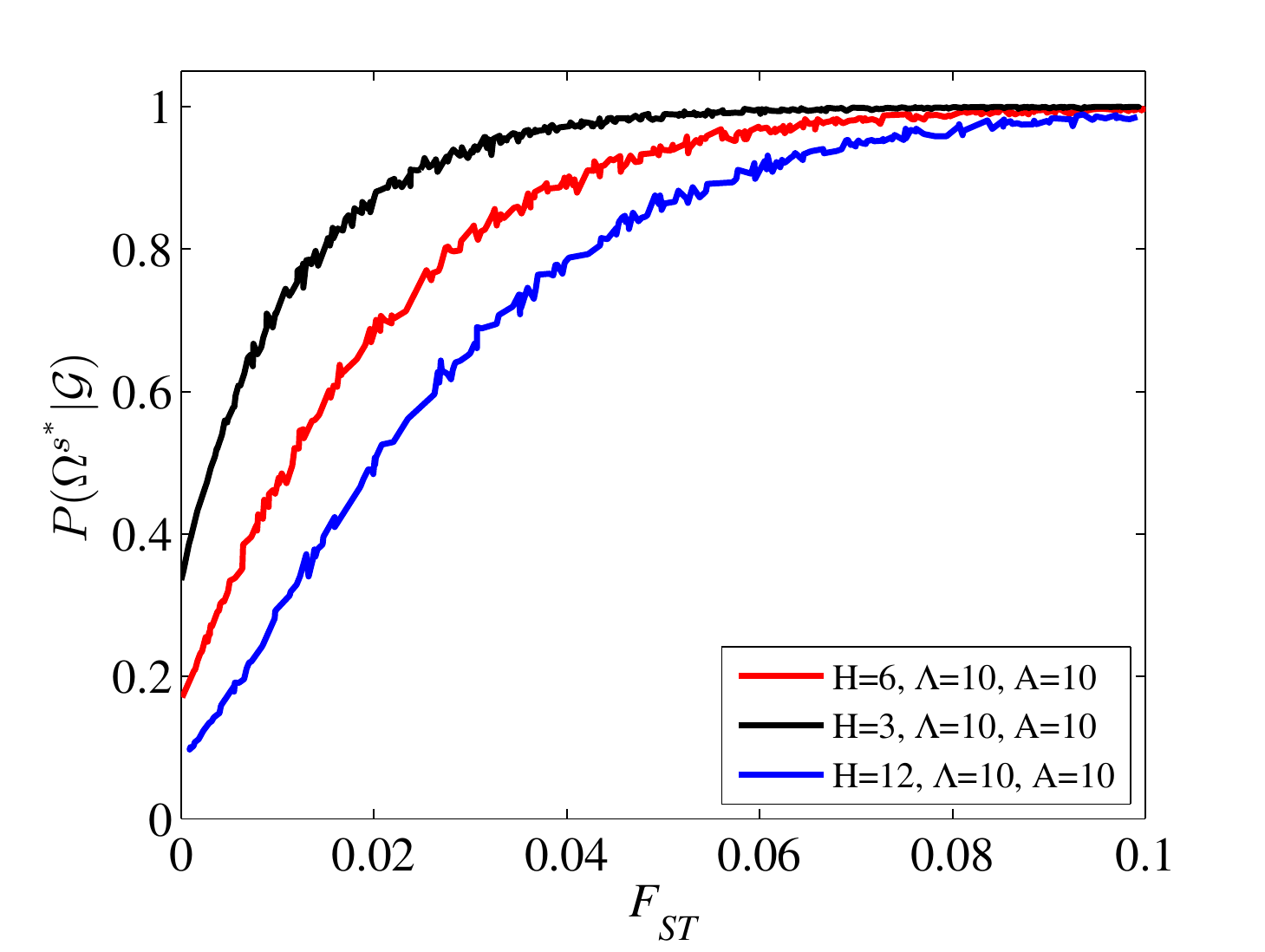}
\caption{Discrimination power: average posterior probability that an individual comes from its true habitat, given its genotype $\mathcal G$, in terms of the $F_{ST}$. Left: effect of the number of loci $\Lambda$ and of the number of alleles per locus $A$. Right: effect of the number of habitats $H$. In both figures the red curve corresponds to the parameters that have been used in this study.}
\label{fig:FST}
\end{figure}

To better quantify the power of our genetic system to discriminate between different habitats of origin, depending on the $F_{ST}$, we computed the average posterior probability that an individual comes from its true habitat, say $\Omega^{h^*}$, given its genotype $\mathcal{G}$, that is:
$\mathds{P}(i\hbox{ comes from }\Omega^{h^*}|\mathcal{G})$. Using Bayes theorem and considering that all of the $H$ habitats are equally likely a priori, we get:
\be
\mathds{P}(i\hbox{ comes from }\Omega^{h^*}|\mathcal{G})=\frac{\mathds{P}(\mathcal{G}| i\hbox{ comes from }\Omega^{h^*})/H}{\sum\limits_{h=1}^H \mathds{P}(\mathcal{G}| i\hbox{ comes from }\Omega^{h})/H}.
\ee
For 300 $F_{ST}$ values between $0$ and $0.1,$ we simulated $1000$ data sets (allele frequencies in the $H$ sites, see Section~\ref{sec:simu_data}), we sampled $1000$ genotypes $\mathcal{G}$ in one of the sites, and we averaged the quantity $\mathds{P}(i\hbox{ comes from }\Omega^{h^*}|\mathcal{G})$ over the 1000 individuals. We call the obtained quantity the \textit{discrimination power} of our genetic system.
As expected, increasing the $F_{ST}$ leads to higher discrimination power (Fig.~\ref{fig:FST}).  With our parameters ($H=6$, $\Lambda=10$, $A=10$), the discrimination power is about 0.5 for an $F_{ST}$ of 0.01; it reaches 0.9 for an $F_{ST}$ of 0.05 and 0.99 for an $F_{ST}$ of 0.1.
This explains the gap between the quality of the estimators obtained with the $F_{ST}$ values larger than $0.05$, compared to the case $F_{ST}=0.01.$ This could also explain the small difference in the accuracy of the estimation when the $F_{ST}$ is increased to $0.1,$ compared to the case $F_{ST}=0.05.$ Additionally, we note that it does not seem necessary to go beyond the value  $F_{ST}$=0.1 to get a very high discrimination power. This also means that the variability in our estimator $\hat D,$ when $F_{ST}$=0.1, is not due to an uncertainty on the habitat of origin of the genotyped individuals, but rather to the sampling variability due to the limited number of these individuals.

Fig.~\ref{fig:FST} also shows the effect of the number of habitats, the number of loci and the number of alleles. Increasing the number of alleles per locus or the number of loci have a comparable effect, with in both cases an increase in the discrimination power. For $F_{ST}$ values close to 0,  the discrimination power converges to $1/H,$ meaning that very small genetic differentiation between the subpopulations leads to equiprobable habitat of origin. In such case, the data contain no information on the origin of the trapped individuals, and the estimation of $D$ is therefore not possible.  For larger values of the $F_{ST}$, increasing the number of sites always leads to lower discrimination power.

\begin{figure}
\centering
\includegraphics[width=0.5\textwidth]{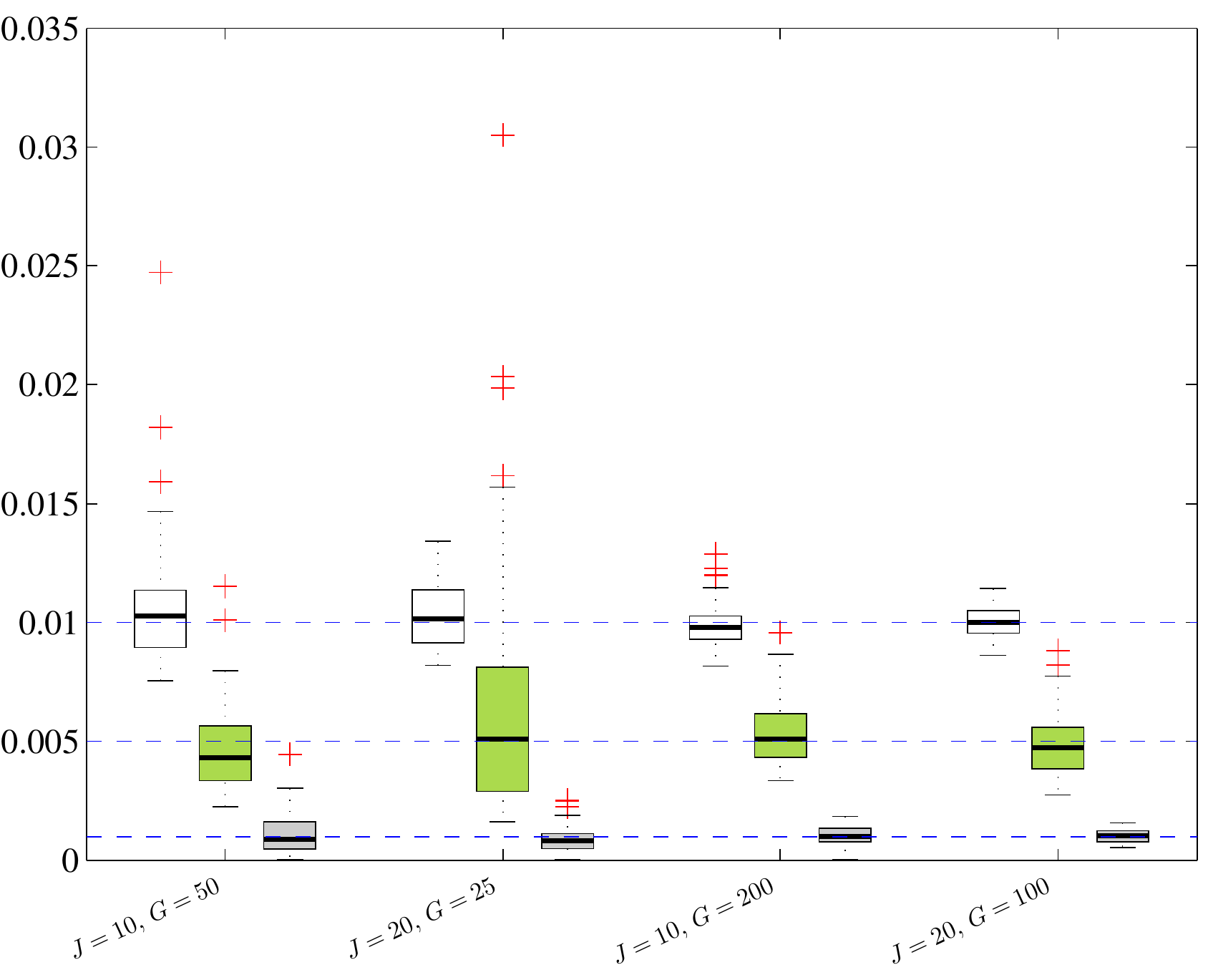}

 \caption{Effect of number of traps and of the total number of genotyped individuals on the quality of the estimator $\hat D(x)$ in each of the three regions: $\hat D_1$ in the matrix (in white), $\hat D_2$ in the habitats (in green) and $\hat D_3$ in the barrier (in grey). The blue dashed lines correspond to the true values of $D$ in the three regions: $D_1=0.01,$  $D_2=0.005$ and $D_3=0.001.$}
 \label{fig:bp2}
\end{figure}

The results in Fig.~\ref{fig:bp2} show, for $F_{ST}=0.1,$ the effect of the number of traps and of the total number of genotyped individuals on the quality of the estimator. In all cases, the median of $\hat D$ is close to the true value, in each of the three regions. For a fixed total number of genotyped individuals, increasing the number of traps $J$ from $10$ to $20$ does not increase significantly the quality of the estimator. It can even decrease it in some situations, e.g., the interquartile range is about twice larger with $20$ traps than with $10$ traps in the habitats. A possible explanation for this counterintuitive effect is that, when the number of traps is increased, the number $G$ of individuals genotyped per trap decreases; in the regions where no traps are added (here, the habitats, see Fig.~\ref{fig:land}), this can lead to more uncertainty on the estimator.

The results of Fig.~\ref{fig:bp2} confirm that increasing the number of genotyped individuals leads to far better estimations: in both cases $J=10$ and $J=20,$ the interquartile ranges are divided by more than two in average when the total number of genotyped individuals is increased from $J\times G =500$ to $J\times G=2000,$ and the distance of the outliers to the median (i.e., approximatively to the true value) is reduced.

\section{Discussion \label{sec:Discussion}}

Using broadly-recognized population models based on a mechanistic description of individual movements \citep{PreAge04,SmoFoc10}, we have developed an approach to estimate the local effect of the environment on individual mobility, based on genetic data. In an environment made of three types of regions, each one associated with a different level of mobility -- or diffusion -- we successfully estimated the diffusion parameters $D(x)$ in each region. The reaction-diffusion framework enabled a fast computation of expected population densities, making parameter estimation possible in a reasonable time. Genetic data had already proved their effectiveness in the estimation of dispersal kernels \citep{RobGar07,KleBon13} in more empirical models. Our results show that successful estimation of parameters of mechanistic population models is also possible using genotype measurements during dispersal and allele frequency data before dispersal, without the need of abundance or mark-recapture data. Genetic data lead to likelihoods of arriving from a given source, and are insensitive to the global population size (i.e., to the parameter $\alpha$ in our approach) and to the relative efficiency of the different traps (the parameters $\beta_\tau$). This advantage of working on probabilities of originating from the different sources has already been shown for kernel estimations \citep{RobGar07,KleBon13}.
A related approach, proposed by \cite{OvaRek08} allowed to estimate the parameters of a diffusion model, based on mark-recapture data, with a single type of marks. In our framework, the genotype information, given the allele frequencies in the different habitats constituting our study-site, can be seen as mark-recapture data, with several types of marks (one per habitat) and some uncertainty on the marks of the captured individuals. Mark-recapture experiments with several types of marks should lead to good estimation results, as they would combine the advantages of our method (insensitivity to several parameters) and of traditional mark-recapture experiments (no unknown external sources, perfect knowledge of the frequencies).

The genetic differentiation between the subpopulations corresponding to the different habitats of origin plays a key role in the quality of the estimation, as shown by the strong effect of the $F_{ST},$ especially on the variability of the estimator. Low $F_{ST}$ values are associated with larger standard deviations and interquartile ranges, which can be explained by a lower posterior probability associated to the true habitat of origin of a genotyped individual. Defining the discrimination power of our genetic system as the average of this posterior probability, we could disentangle the effect of the $F_{ST}$ and that of the other sources of uncertainty in our estimator of the diffusion parameters $D(x).$ With an intermediate level of genetic differentiation ($F_{ST}=0.05$), the discrimination power was high (0.9 with 20 traps and 2000 genotyped individuals), and the quality of our estimator of $D(x)$ was comparable to the case $F_{ST}=0.1$. With an $F_{ST}$ of $0.1,$ the discrimination power was close to $1$, which is almost exact; the remaining uncertainty in the determination of $D(x)$ may therefore be sampling variance due to the finite number of genotyped individuals per trap. It could also be due to the lack of uniqueness in the inverse problem of determining $D(x),$ even with infinite population sizes. From a theoretical viewpoint, the unique determination of diffusion and conductivity coefficients based on a finite set of measurements is a difficult problem, as illustrated by the Calder\'{o}n problem \citep{Cal80} of determining the electrical conductivity of a medium for which uniqueness is only proved with infinitely many observations~\citep{SylUhl87,Nac96}.

As expected, the size of the post-dispersal sample has a clear effect on the uncertainty of the estimation. The quality of the estimation, and especially its variability, is clearly improved when the number of genotyped individuals is increased. If the total number of genotyped loci and the number of traps were fixed,  the trade-off between increasing the number of loci and the number of genotyped individuals per trap would depend on the main source of uncertainty: increasing the number of loci per individual increases the discrimination power while increasing the number of genotyped individuals per trap reduces the sampling variance. The role of the number of traps is less obvious. Intuitively, increasing the number of traps leads to a better coverage of the study site, which should have a positive impact on the estimation. However, with a fixed total number of genotyped individuals, this leads to a decrease in the number of genotyped individuals per trap and can therefore produce more uncertainty in the regions where the number of traps has not been increased (the habitats in our simulations). It should be noted however that too few traps may lead to identifiability problems, as would in the extreme case of a unique trap placed at equidistance between two habitats in an otherwise homogeneous landscape.

Remarkably, the estimation of $D(x)$ in the matrix remains accurate, with no outliers among the estimators of $D(x)$  even with low $F_{ST}$ values. This may be the consequence of the larger area of the matrix compared to the other regions in the landscape leading to a stronger effect of the value of the diffusion parameter in this region. Conversely, larger standard deviations and interquartile ranges are observed in the habitats, for all $F_{ST}$ values. This cannot be fully explained by the smaller area of the habitat region, as the estimation on the barrier is more accurate, with a comparable area. Most of the individuals trapped in a given habitat come from the same habitat (about $90\%$). The remaining individuals being sparse, this leads to higher relative variance in the proportions of individuals trapped in the habitats than in the other regions, which can explain the lower accuracy of the estimation of the diffusion parameter in the habitats. Based on these observations, we suspect that the estimation of a single coefficient in a homogeneous environment would most likely be reliable, even with low $F_{ST}$ values, and that placing the traps far from the release sites should lead to a better estimation of the coefficient in such case.

In addition to the Hardy-Weinberg equilibrium and independence of loci, an important assumption in our approach was that the allele frequencies were exactly known in the habitats. In practice, the frequencies are determined from previously sampled populations. The sample size is known to have an important effect on successful assignment of genotyped individuals \citep{CorPir99}. Reducing this size should lead to some uncertainty in the allele frequencies with an effect comparable to that of decreasing the $F_{ST}$, i.e, lowering the discrimination power of the genotype data. A problematic case noted in \cite{PaeCal95} and \cite{CorPir99} while studying assignment methods is when some individuals carry an allele which has not been detected in the sample corresponding to their population of origin, leading to a null posterior probability that these individuals come from their true habitat. In such case, \cite{PaeCal95} suggest to add the genotype of these individuals to the population samples defining the allele frequencies in all of the habitats.

In our study, the allele frequencies are determined before dispersal, in individuals from the same generation as the trapped individuals. At each generation, the allele frequencies are modified, due to drift and gene flow among habitat patches. Thus, using allele frequencies sampled from previous generations could lead to an inaccuracy in the frequencies which depends on the gene flow and on the number of generations before the capture of the genotyped dispersers. To estimate the effect of the population flows on the allele frequencies after one generation, we computed the quantity $C_\tau^h/C_\tau,$ for $\tau=h,$ corresponding to the proportion of individuals captured in $\Omega^h,$ whose habitat of origin is $\Omega^h$. The values clearly depend on the proximity of another habitat, with a ratio of $0.98$ in $\Omega^1,$ which is the farthest from the other habitats, and of $0.82$ in $\Omega^3$ and $\Omega^5$ which are close to each other (the other ratios are $0.87$, $0.85$ and $0.94$ in $\Omega_2,$ $\Omega_5$ and $\Omega_6,$ respectively). Thus, the allele frequencies may remain stable after several generations if the habitats are sufficiently far from each other or become rapidly inaccurate in the opposite situation. A difficulty in estimating whether two habitats are far from each other is that the population flows are not known a priori.

Our approach was based on an unbiased Brownian motion description of the individual trajectories, which leads to a Fokker-Planck reaction-diffusion equation. It can easily be extended to include a bias in a direction $(v_1,v_2)(t,x)$ modelling attractiveness or repulsiveness of some elements of the landscape, by adding a term $-\partial (u \, v_1)/\partial x_1 -\partial (u \, v_2)/\partial x_2$ in the reaction-diffusion equation \eqref{eq:RDu}. The method proposed here also readily applies to other linear dispersal terms, such as the Fickian diffusion $\nabla \cdot (D(x) \nabla u)$ which is nevertheless more adapted to describe electric and thermic conductivity \citep{RoqAug08,Roq13}. More general L\'evy processes than Brownian diffusion, corresponding to movements with large jumps, could have been considered as well; in such cases, the Laplace diffusion operator $\Delta$ would be replaced by a fractional Laplace operator \citep{Val09}. Integral kernel-based dispersal terms which can account for long distance dispersal events could also be considered in place of the diffusion approach \citep{KotLew96}.

The purpose of our study was to propose a rigourous method for the estimation of dispersal parameters in stochastic differential equation-based models of individual movement. Here, the method performance analysis assumed the same type of model for simulations and for inference. In practice, the real dispersal process may differ from the assumptions of the model. Future work should focus on the robustness of the method when the model assumptions are violated.

Another possible extension of our study is to include the estimation of the relative pre-dispersal densities in the different habitats of origin and/or the areas of these habitats. It is not straightforward however that these problems are identifiable.

As suggested by Remark~\ref{rem:q_vs_c}, pre-dispersal subpopulations may not coincide with the habitats. The location of the subpopulations may also not be known a priori. A challenging extension of our approach would consist in clustering the subpopulations and inferring the allelic frequencies \citep[following a Structure-like approach, see][]{PriSte00} together with the estimation of $D(x)$ in a full Bayesian approach, based on pre-dispersal genotype data in the habitats and genotype data of trapped dispersers.


\section*{Appendix~A: gradual release of the pre-dispersal populations}

The equation \eqref{eq:RDu} describes a simultaneous release of all the individuals at $t=0.$ To account for a possible gradual release of the individuals, the equation~\eqref{eq:RDu} can be replaced by:
\be \label{eq:RDu_f}
\frac{\partial u}{\partial t}=\Delta (D(x) \, u) -\frac{u}{\nu}+u_0(x)\, f(t), \ t>0, \, x\in \Omega,
\ee
where the term $u_0(x)\, f(t)$ describes the release of the individuals;  $u_0(x)$ still corresponds to the pre-dispersal density and the function $f(t)$ is the release rate. It can be described by any nonnegative function or distribution with integral $1$ and with support in $[0,T],$ $T$ corresponding to the end of the release period. In this framework, the  density of dispersers coming from habitat $\Omega^h$ satisfies the equation:
\be \label{eq:RDuc_f}
\frac{\partial u^h}{\partial t}=\Delta (D(x) \, u^h) -\frac{u^h}{\nu}+u_0^h(x)\, f(t), \ t>0, \, x\in \Omega,
\ee
where $u_0^h$ is still given by \eqref{eq:predis_h}.

\section*{Appendix~B: precise shape of the diffusion terms}
\begin{figure}[h!]
\centering
\includegraphics[width=0.4\textwidth]{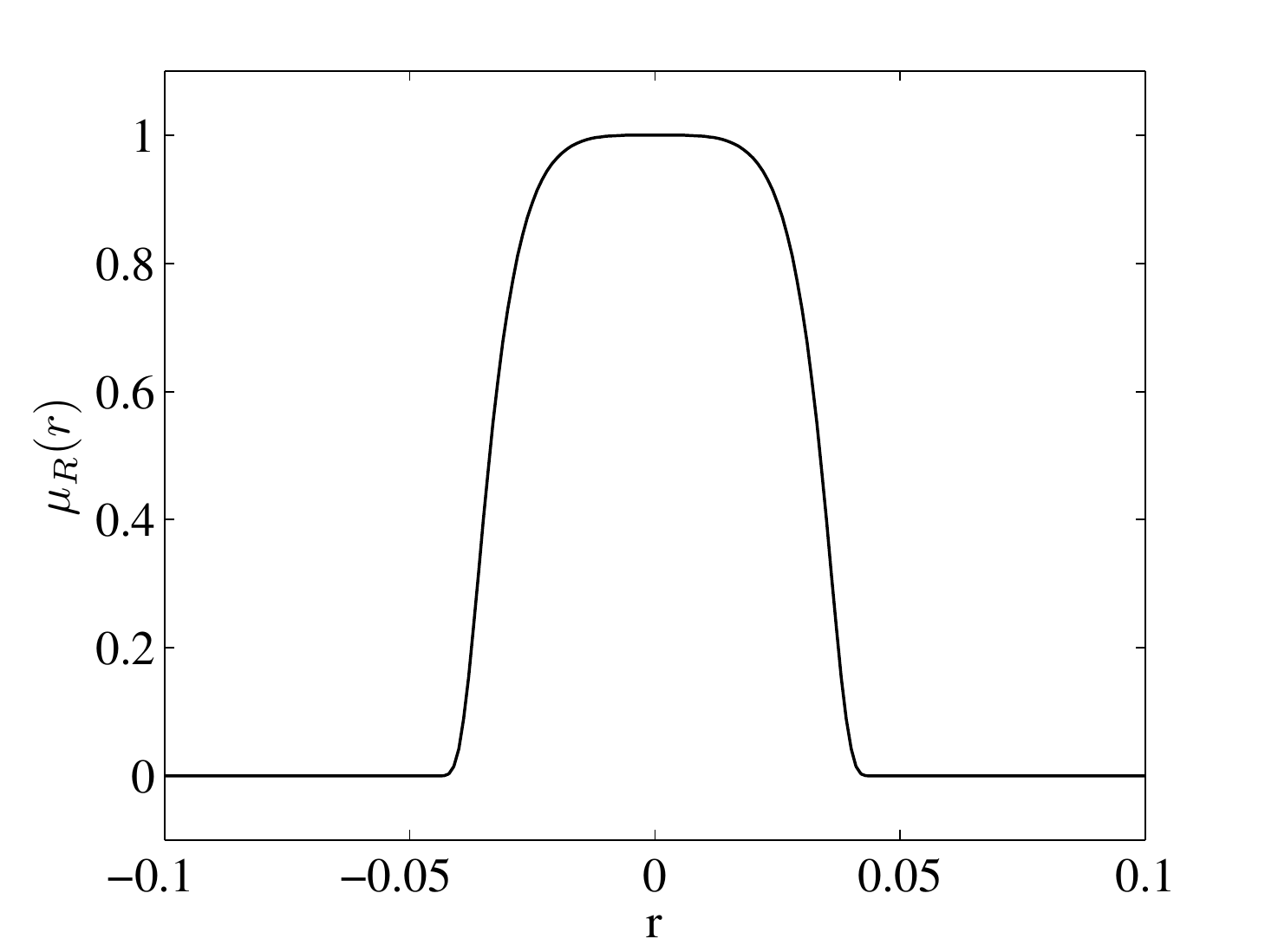}
\caption{The function $\mu_R(r)$, for $R=0.05$ and $r\in (-0.1, 0.1)$.}
\label{fig:muR}
\end{figure}

In our numerical computations, we took
$$\phi(x)=\mu_{2\, R}(\|x\|) \hbox{ and }\psi(x)=\psi(x_1,x_2)=\mu_{R}\lp  x_1- q\rp,$$for the function
$\mu$ defined by (see Fig.~\ref{fig:muR}): $$\mu_R(r)=\exp\left(\frac{-r^4}{(r^2-R^2)^2}\right) \hbox{ for }r \in (-R,R) \hbox{ and }\mu_R(r)=0 \hbox{ otherwize}.$$

\section*{Appendix~C: computation of the $F_{ST}$}

The index $F_{ST}$ is used as a measure of genetic differentiation among the subpopulations. It was computed as follows: we set $$H_S=\frac{1}{\Lambda}\sum\limits_{\lambda=1}^{\Lambda} \sum\limits_{a=1}^{A} \sum\limits_{h=1}^{H} \frac{1}{H}\left(p_{h \lambda a}\right)^2  \hbox{ and } H_T=\frac{1}{\Lambda}\sum\limits_{\lambda=1}^{\Lambda}\sum\limits_{a=1}^{A} \left(\frac{1}{H} \sum\limits_{h=1}^{H} p_{h \lambda a}\right)^2, $$
where $\Lambda$ is the number of loci, $A,$ the number of alleles per locus whose frequency is measured and $H$ the number of subpopulations, and \be F_{ST}= \frac{H_S-H_T}{1-H_T}.\ee This formula corresponds to Nei's $G_{ST}$ for a single locus \citep{Nei73}, with numerator and denominator averaged over the $\Lambda$ loci. In our computations, all the subpopulations had the same size; in other situations, the weight $1/H$ in the above formulas for $H_S$ and $H_T$ should be replaced by the relative sizes of the subpopulations.

\section*{Appendix~D: numerical computation of the cumulated population densities}

In order to compute the cumulated densities $w_\infty(x)$ and $w_\infty^h(x),$  we used the time-dependent partial differential equation solver Comsol Multiphysics$^{\copyright}$ applied to the evolution equations  \eqref{eq:RDw} and \eqref{eq:RDwc} below at large time ($t=20$),  with default parameter values (finite element method with second order basis elements) and a triangular mesh adapted to the geometry of our landscape and made of 5296 elements.

We defined the cumulated population density at intermediate times $t$ and position $x$ by:
\be
w_t(x)=\int_0^{t}u(s,x)\, ds, \ \hbox{ for all }t>0, \ x\in \Omega.
\ee
Integrating \eqref{eq:RDu} between $0$ and $t>0$ we note that  $w_t(x)$ satisfies the following equation:
\be \label{eq:RDw}
\frac{\partial w_t}{\partial t}=\Delta (D(x) \, w_t) -\frac{w_t}{\nu}+u_0(x), \ t>0, \, x\in \Omega,
\ee
and $w_0(x)=0.$

Similarly, the cumulated population density of individuals coming from $\Omega^h$ is:
\be
w_t^h(x)=\int_0^{t}u^h(s,x)\, ds, \ \hbox{ for all }t>0, \ x\in \Omega.
\ee
This function satisfies:
\be \label{eq:RDwc}
\frac{\partial w_t^h}{\partial t}=\Delta (D(x) \, w_t^h) -\frac{w_t^h}{\nu}+u_0^h(x), \ t>0, \, x\in \Omega,
\ee
and $w_0^h(x)=0.$

\

\noindent {\bf Conflict of Interest:} The authors declare that they have no conflict of interest.

%
%

%
%
%
%
%

\end{document}